\documentclass[aps,prd,twocolumn,tightenlines,nofootinbib,superscriptaddress,floatfix]{revtex4-2}

\usepackage{feynmf}
\usepackage{soul}
\usepackage{mciteplus}
\usepackage[utf8]{inputenc}
\usepackage[T1]{fontenc}
\usepackage{lmodern}
\usepackage{microtype}
\usepackage[english]{babel}
\usepackage{amssymb,amsbsy,amsmath,amsfonts,amsthm}
\usepackage{yhmath} 
\usepackage{slashed}
\usepackage{bm}
\usepackage{times}
\usepackage[dvipsnames,table,xcdraw]{xcolor}
\usepackage{graphicx}
\usepackage{epstopdf}
\usepackage{tabularx}
\usepackage{multirow}
\usepackage{longtable}
\usepackage{hyperref}
\usepackage{booktabs}
\hypersetup{colorlinks=true, linkcolor=blue, 
citecolor=cyan, urlcolor=black}
\usepackage{url}
\newcommand{\mathleft}{\@fleqntrue\@mathmargin0pt}
\newcommand{\mathcenter}{\@fleqnfalse}
\usepackage{soul}
\usepackage{orcidlink}
\allowdisplaybreaks

\begin{document}

\title{Simultaneous Dalitz-plot decomposition of the $e^+ e^- \to J/\psi \, \pi \, \pi \, (K \bar{K})$ processes in the 4.13--4.36 GeV region using dispersive final-state interactions}
\author{Viktoriia Ermolina\orcidlink{0009-0005-6965-4840}}
\email[]{vermolin@uni-mainz.de}
\affiliation{Institut f\"ur Kernphysik \& PRISMA$^{++}$ Cluster of Excellence, Johannes Gutenberg Universit\"at,  D-55099 Mainz, Germany}

\author{Igor Danilkin\orcidlink{0000-0001-8950-0770}}
\affiliation{Institut f\"ur Kernphysik \& PRISMA$^{++}$ Cluster of Excellence, Johannes Gutenberg Universit\"at,  D-55099 Mainz, Germany}

\author{Marc Vanderhaeghen\orcidlink{0000-0003-2363-5124}}
\affiliation{Institut f\"ur Kernphysik \& PRISMA$^{++}$ Cluster of Excellence, Johannes Gutenberg Universit\"at,  D-55099 Mainz, Germany}

\date{\today}

\begin{abstract}
We present a joint analysis of the processes $e^+e^- \to J/\psi\pi^+\pi^-$ and $e^+e^- \to J/\psi K^+K^-$ at center-of-mass energies from 4.13 to 4.36 GeV. The amplitudes are constructed using the Dalitz-plot decomposition formalism, with the $e^+e^-$ energy dependence encoded through the $Y(4220)$ and $Y(4320)$ resonant structures together with a nonresonant production mechanism. The scalar $\pi\pi/K\bar K$ final-state interaction is treated dispersively using a coupled-channel Omn\`es representation. This allows us to describe the measured total cross sections and one-dimensional invariant-mass distributions with a single set of energy-independent parameters. We find that a purely resonant description of the BESIII data is insufficient, requiring a nonresonant term at the amplitude level which undergoes $\pi\pi/K\bar{K}$ rescattering. Within the present isobar model, we extract Breit-Wigner parameters for the $Z_c(3900)$, $Y(4220)$, and $Y(4320)$ states, and determine the corresponding subprocess cross sections.
\end{abstract}
\maketitle

\section{Introduction}\label{intro} 

Over the past few decades, a large number of states containing heavy charm or bottom quarks have been observed that do not fit naturally into the conventional quark-antiquark meson spectrum \cite{Brambilla:2019esw, Chen:2016qju, Chen:2022asf, Olsen:2017bmm, Karliner:2017qhf, Guo:2017jvc, Liu:2019zoy}. The $e^+e^-$ annihilation process provides a particularly clean environment for studying such states and has played a central role in many discoveries by the \textit{BABAR}, Belle/Belle II, and BESIII experiments
\cite{BaBar:2005hhc, Belle:2007dxy, BaBar:2012vyb, Belle:2013yex,
BaBar:2006ait, BaBar:2012hpr, Belle:2007umv, Belle:2014wyt,
BESIII:2016bnd, BESIII:2020oph, BESIII:2022kcv, BESIII:2022joj,
BESIII:2017tqk, BESIII:2021njb, BESIII:2016adj, BESIII:2014rja,
BESIII:2019gjc, BESIII:2018iea, BESIII:2020bgb, BESIII:2023tll}. Among the most prominent charged charmoniumlike candidates are the $Z_c(3900)$ and $Z_c(4020)$ states \cite{BESIII:2020oph, BESIII:2026ret, BESIII:2025qkn, BESIII:2013ris, Belle:2013yex, BESIII:2013ouc}. They are observed in hidden-charm final states produced in the energy region of the vector structures $Y(4220)$ and $Y(4320)$ \cite{BESIII:2022qal, BESIII:2022joj, BESIII:2025bce, BESIII:2021njb, BESIII:2026qhe, BaBar:2005hhc, Belle:2007dxy, BESIII:2016bnd}, making the combined study of the $Y$ and $Z_c$ sectors an important step toward understanding the dynamics of these exotic candidates.

Several analyses have addressed individual invariant-mass distributions in the processes $e^+e^- \to J/\psi\pi^+\pi^-$ and $e^+e^- \to J/\psi K\bar K$ \cite{BESIII:2017bua, BESIII:2026ret, BESIII:2025qkn, BESIII:2020oph, Danilkin:2020kce}. However, a simultaneous description of the invariant-mass distributions and the energy-dependent total cross sections remains challenging, in particular, connecting the $Z_c$ structures seen in the $J/\psi\pi$ invariant-mass spectra with the $Y$ structures observed in the total cross sections. Moreover, the final states are not produced exclusively through intermediate $Y$ resonances, as indicated by the measured cross sections \cite{BESIII:2022qal, BESIII:2022joj}. The corresponding nonresonant production mechanism can therefore play an important role in the one-dimensional invariant-mass distributions \cite{Nakamura:2023obk}. 

In this paper, we construct an amplitude framework in which the $e^+e^-$ energy dependence is incorporated explicitly. The amplitudes are based on the Dalitz-plot decomposition formalism \cite{JPAC:2019ufm} and include the $Y(4220)$ and $Y(4320)$ resonant contributions together with a nonresonant production term. The scalar $\pi\pi/K\bar K$ final-state interaction is treated dispersively through a coupled-channel Omn\`es representation \cite{Ermolina:2024uln}. This allows us to describe the processes $e^+e^- \to J/\psi\pi^+\pi^-$ and $e^+e^- \to J/\psi K^+K^-$ over the measured energy range with a single set of energy-independent parameters. The same framework also provides predictions for invariant-mass distributions at energies where no such measurements are currently available.

The paper is organized as follows. In Sec.~\ref{sec2}, we present the formalism. In Sec.~\ref{sec3}, we apply the framework to the BESIII invariant-mass distributions \cite{BESIII:2022joj, BESIII:2025qkn} and to the measured total cross sections \cite{BESIII:2022qal, BESIII:2022joj}. We then discuss the resulting resonance parameters, subprocess cross sections, and predictions for unmeasured invariant-mass distributions. The summary and conclusions are given in Sec.~\ref{sec4}.

\section{Formalism}
\label{sec2}

We consider the processes
$e^+e^- \to \gamma^\ast \to J/\psi h^+h^-$, with $h=\pi,K$,
and denote the center-of-mass (c.m.) energy by $\sqrt{s}$. For the
$J/\psi\pi^+\pi^-$ channel, we label the final-state particles as
\begin{align}\label{eq:labels_1}
&1=\pi^- ,\qquad 2=J/\psi ,\qquad 3=\pi^+.
\end{align}
For the $J/\psi K^+K^-$ channel, we use the analogous convention
\begin{align}
&1=K^- ,\qquad 2=J/\psi ,\qquad 3=K^+\,.\nonumber
\end{align}
The Mandelstam variables are defined as
\begin{align}
\sigma_k = (p_i+p_j)^2 \equiv m_{ij}^2,
\quad
(ij)k\in\{(23)1,(31)2,(12)3\}\,,\nonumber
\end{align}
which satisfy
\begin{align}
\sigma_1+\sigma_2+\sigma_3
=
s+m_1^2+m_2^2+m_3^2 .
\label{eq:sigma_sum}
\end{align}
The QED vertex allows one to write the differential cross section as
\begin{align}\label{eq:cross_section}
   & \frac{d\sigma}{d\phi_1 d\cos{\theta_1} \, d\phi_{23} \, d\sigma_1 \, d\sigma_2}=\frac{e^2}{64\,(2\pi)^{5}\,s^3}  \\
   & \quad\quad\quad \times \sum_{\{\lambda\}} \left(\left| M^{\Lambda=+1}_{\{\lambda\}} \right|^2+\frac{2m^2_e}{s}\left| M^{\Lambda=0}_{\{\lambda\}} \right|^2\right),\nonumber
\end{align}
where $m_e$ is the electron mass, $\Lambda$ denotes the virtual-photon helicity, and $\{\lambda\}$ collectively denotes the final-state helicities. In the present case, since the two light mesons are pseudoscalars, this set reduces to the $J/\psi$ helicity. In the following, we neglect the contribution of the helicity amplitude associated with the helicity $\Lambda=0$ of the virtual photon, as it is suppressed by a factor of $2m^2_e/s$ for  $s \gg m_e^2$. The angles $(\phi_1,\theta_1,\phi_{23})$ provide the orientation of the decay plane in the $e^+e^-$ c.m. frame.

\subsection{Dalitz-plot decomposition}

To define the transition helicity amplitude, we employ the recently proposed Dalitz-plot decomposition (DPD) factorization \cite{JPAC:2019ufm}. This formalism conveniently describes many-body final states and isolates the specific spin and parity $J^P$ contributions (isobars) that arise in the $\gamma^* \to J/\psi \, \pi \, \pi \, (K \bar{K})$ process. Furthermore, the DPD approach has been shown to have an exact mapping from a Lagrangian formulation \cite{Ermolina:2024uln}. This construction allows to isolate the decay plane's orientation and express the remaining part in terms of Mandelstam variables
\begin{equation}\label{eq:amp}
    M^\Lambda_{\{\lambda\}} =\sum_{\nu} D^{J*}_{\Lambda,\nu}(\phi_1,\theta_1,\phi_{23})\,O^\nu_{\{\lambda\}}(\{\sigma\})\,,
\end{equation}
where the virtual photon has spin $J$ and the rotation with Wigner $D$-function connects the frame of calculation with the actual $e^+e^-$ c.m. frame. The angle $\phi_1$ is a global azimuthal angle. Since the initial $e^+e^-$ state is azimuthally symmetric, it is redundant for the unpolarized observables considered here and gives a factor $2\pi$ after integration. Further details on the amplitude construction can also be found in \cite{Ermolina:2024uln}.

The Dalitz-plot function $O^\nu_{\{\lambda\}}$ is given by a product of individual two-particle decays occurring in between, each considered in the rest frame of a decaying particle
\begin{equation}\label{eq:DPD_O_general}
\begin{split}
    O^\nu_{\{\lambda\}}(\{\sigma\}) & = \sum_{(ij)k} \sum_s^{(ij)\to i,j} \sum_\tau \sum_{\{\lambda'\}} n_J\, n_s\,d^J_{\nu,\tau-\lambda'_k} (\hat{\theta}_{k(1)}) \\& \times \,H^{0 \to (ij),k}_{\tau,\lambda'_k}X_s (\sigma_k)\,d^s_{\tau,\lambda'_i-\lambda'_j}(\theta_{ij})\,H^{(ij) \to i,j}_{\lambda'_i,\lambda'_j} \\& \times d^{j_1}_{\lambda'_1,\lambda_1} (\zeta^1_{k(0)})\, d^{j_2}_{\lambda'_2,\lambda_2} (\zeta^2_{k(0)})\, d^{j_3}_{\lambda'_3,\lambda_3} (\zeta^3_{k(0)})\,.
\end{split}
\end{equation}
Here, the first sum runs over all possible configurations $(ij)k \in \{(23)1,(31)2,(12)3\}$, which correspond to three potential decay chains. The second and third summations run over the various possible spins $s$ and helicity $\tau$ of the isobar $(ij)$. Individual spins of the final-state particles are denoted by $j_i$. Rotations by the angles $\hat{\theta}_{k(1)}$ relate the three decay chains one to another, which is achieved by choosing a specific calculation frame, while the angles $\theta_{ij}$ denote the polar angle of a particle $i$ in the $(ij)$ rest frame. Finally, a boost that induces the rotation of the final-state helicities by the angles $\zeta^{1,2,3}_{k(0)}$ connects the individual two-body decays. Detailed expressions for angles $\hat{\theta}_{k(1)},\theta_{ij},\zeta^{1,2,3}_{k(0)}$ in terms of Mandelstam variables can be found in Appendix A of~\cite{JPAC:2019ufm}. Finally, the functions $H$ denote the helicity couplings, $X_s (\sigma)$ specifies the subchannel line shape, and $n_J$ and $n_s$ serve as conventional normalization factors.

The factorization itself is essentially model independent. However, the helicity couplings are generally unknown and must be determined by fitting to the available data. For this purpose, it is convenient to employ the $LS$-coupling scheme \cite{Chung:1997jn}. For a particle $0$ that decays into an isobar $(ij)$ and a final-state particle $k$, as well as for the decay of the isobar $(ij)$ into particles $i$ and $j$, the decompositions are given by
\begin{alignat}{2}
&H^{0 \to (ij),k}_{\tau,\lambda'_k} 
  &&= \sum_{LS} \alpha^{0 \to (ij),k}_{LS} \sqrt{\frac{2L+1}{2J+1}} \langle s,\tau;j_k,-\lambda'_k|S,\tau-\lambda'_k \rangle \nonumber \\
  &&&\times \langle L,0;S,\tau-\lambda'_k|J,\tau-\lambda'_k \rangle p^L \,B_L\,, \nonumber \\
&H^{(ij) \to i,j}_{\lambda'_i,\lambda'_j} 
  &&= \sum_{l's'} \alpha^{(ij) \to i,j}_{l's'} \sqrt{\frac{2l'+1}{2s+1}} \langle j_i,\lambda'_i;j_j,-\lambda'_j|s',\lambda'_i-\lambda'_j \rangle \nonumber \\
  &&&\times \langle l',0;s',\lambda'_i-\lambda'_j|s,\lambda'_i-\lambda'_j \rangle {p'}^{l'}\, B_{l'}\,. \label{eq:ls}
\end{alignat}
Here, $\alpha^{0 \to (ij),k}_{LS}$ and $\alpha^{(ij) \to i,j}_{l's'}$ denote the $LS$ couplings of the corresponding decays, $S$ is the spin of the isobar-spectator system, $s'$ is the spin of the $i$-$j$ system, and $L, l'$ are the relative orbital angular momenta between the final-state particles.  The magnitude of momenta $\vec{p}_k$ or $\vec{p}_i+\vec{p}_j$ in the rest frame of a particle $0$ is denoted by $p$ \cite{Zou:2002ar}, while $p'$ is the magnitude of $\vec{p}_i$ or $\vec{p}_j$. The quantities $B_L$ and $B_{l'}$ are the Blatt-Weisskopf barrier factors (normalized to $1$ at the resonance position) \cite{VonHippel:1972fg}, which ensure the proper asymptotic behavior.

\begin{table}[t!]
\centering 
\renewcommand*{\arraystretch}{1.4}
\begin{tabular*}{\columnwidth}{@{\extracolsep{\fill}}lc}
  \hline
 Decay & Corresponding $LS$ $(l's')$ combinations \\ 
 \hline
 $\gamma^* \to Z^\pm_c\pi^\mp$ & $(0,1)$, $(2,1)$ \\ 
 $Z^\pm_c \to J/\psi\pi^\pm$ & $(0,1)$, $(2,1)$ \\ 
 $\gamma^* \to f_0 J/\psi$ & $(0,1)$, $(2,1)$ \\
 $f_0 \to \pi^+\pi^-/K^+K^-$ & $(0,0)$ \\
 $\gamma^* \to f_2 J/\psi$ & $(0,1)$, $(2,1)$, $(2,2)$, $(2,3)$, $(4,3)$ \\
 $f_2 \to \pi^+ \pi^-$ & $(2,0)$\\
 \hline
\end{tabular*}
\caption{Allowed $LS$ $(l's')$ combinations in Eq. (\ref{eq:ls}) for each of the two-body decays involved in the $e^+ e^- \to J/\psi \, \pi^+ \, \pi^-$ process. In the nominal fit only the lowest-$L(l')$ entry in each row
is retained.}
\label{tab1}
\end{table}

\begin{figure*}[t]
	\centering 
	\includegraphics[width=0.93\textwidth]{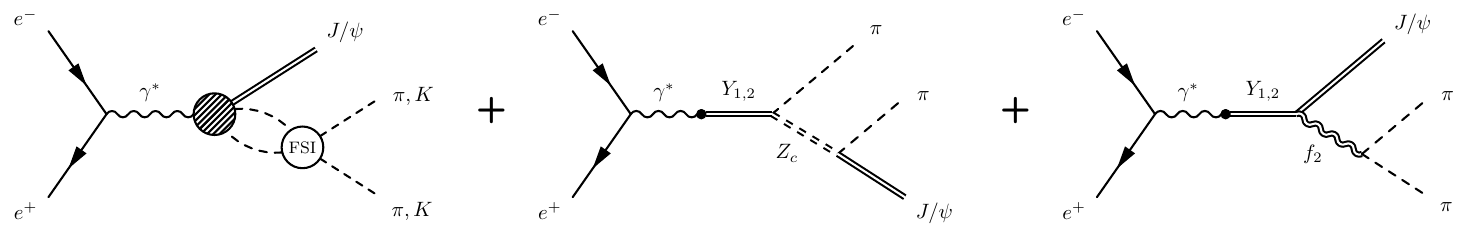} \\
    \vspace{12pt} 
    \includegraphics[width=0.77\textwidth]{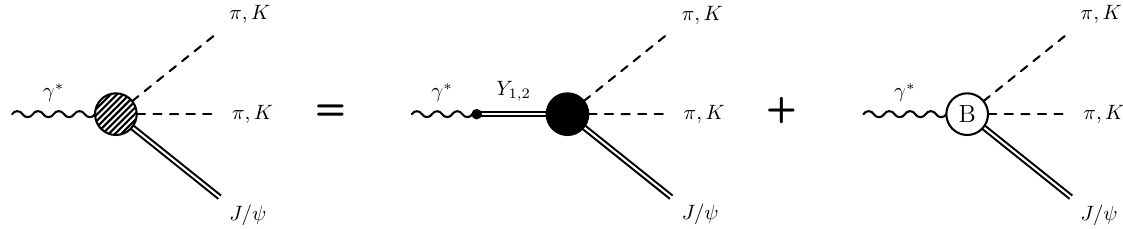}
    \caption{Diagrams illustrating the considered contributions to the $e^+ e^- \to J/\psi \, \pi \, \pi \, (K \bar{K})$ process.}
	\label{diag1}
\end{figure*}

Equations~\eqref{eq:DPD_O_general} and \eqref{eq:ls}
represent the general DPD construction for a three-body decay
$0\to 123$. We now specify this construction for the processes $\gamma^\ast\to J/\psi h^+h^-$, with $h=\pi,K$. In the present case the initial particle is the virtual photon, $J=1$, while the two light mesons are pseudoscalars. Therefore the only nonzero
final-state helicity is that of the $J/\psi$, which is denoted by $\lambda_2$. With our convention \eqref{eq:labels_1}, the three possible DPD chains correspond to
\begin{align}
&(23)1:\quad J/\psi\pi^+ \ {\rm with\ spectator}\ \pi^- ,
\nonumber\\
&(12)3:\quad J/\psi\pi^- \ {\rm with\ spectator}\ \pi^+ , \\
&(31)2:\quad \pi^+\pi^- \ {\rm with\ spectator}\ J/\psi .
\nonumber
\end{align}
In the $J/\psi\pi^\pm$ subsystems, we include the $Z_c^\pm(3900)$ contribution, while in the $\pi^+\pi^-$ subsystem we include the scalar $f_0(500)$, $f_0(980)$ and tensor $f_2(1270)$ contributions. For the $J/\psi K^+K^-$ channel we retain the scalar-isoscalar $K\bar K$ contribution generated by the coupled-channel $\pi\pi/K\bar K$ final-state interaction. The two charged $Z_c$ chains are related by crossing or, equivalently, by interchanging the two pion labels. Within the DPD convention this relation involves the corresponding helicity-frame phase factors~\cite{JPAC:2019ufm, Ermolina:2024uln}, which allow the same $LS$ couplings to be used for the $Z_c^+$ and $Z_c^-$ chains. Table~\ref{tab1} lists the parity-allowed $LS$ combinations for the two-body transitions entering the $J/\psi\pi^+\pi^-$ amplitude. In the nominal fit, only the lowest allowed $LS$
contribution in each channel is retained.

\subsection{Initial energy dependence}

To capture both resonant and nonresonant production mechanisms, we explicitly factorize the $s$ energy dependence out of the vertices, rendering the underlying couplings globally constant. As seen from the total cross section of the $e^+ e^- \to J/\psi \, \pi \, \pi \, (K \bar{K})$ processes \cite{BESIII:2022qal, BESIII:2022joj}, one must include the $Y(4220)$, $Y(4320)$ resonances and, in addition, the nonresonant background (as shown in Fig.~\ref{diag1}).
We denote
\begin{equation}\nonumber
Y_1\equiv Y(4220),\qquad Y_2\equiv Y(4320).
\end{equation}
The resonant part is motivated by vector-meson dominance, in which the
virtual photon couples to intermediate vector states $Y_i$. We use the
Breit-Wigner-type propagator
\begin{equation}\label{eq:prop}
P_{R}(x)=\frac{1}{x-m^2_{R}+im_{R}\Gamma_{R}},
\end{equation}
where $x=s$ for the vector states $Y_i$ and $x=\sigma_k$ for
subchannel resonances such as $Z_c(3900)$ and $f_2(1270)$. With this convention, the DPD building blocks for the two
$Z_c^\pm(3900)$ chains are replaced by
\begin{align}\label{eq:zcprop}
&\alpha^{0 \to (23),1}_{01} X_{1}\,\alpha^{(23) \to 2,3}_{01} \\
  &= P_{Z_c(3900)}(\sigma_1) \Big[\alpha_{1} e^{i\phi_{Z_1}}  P_{Y_1}(s) \nonumber 
  + \alpha_2 e^{i\phi_{Z_2}} P_{Y_2}(s) \Big], \\
&\alpha^{0 \to (12),3}_{01} X_{1}\,\alpha^{(12) \to 1,2}_{01} =(-1)^{1-\lambda'_2} \nonumber \\
  & \times P_{Z_c(3900)}(\sigma_3) \Big[\alpha_1 e^{i\phi_{Z_1}} P_{Y_1}(s) 
  + \alpha_2 e^{i\phi_{Z_2}} P_{Y_2}(s) \Big]\nonumber.
\end{align}
The factor $(-1)^{1-\lambda'_2}$ follows from the DPD helicity-frame
phase convention when the two pion labels are interchanged \cite{Ermolina:2024uln}. The parameters
$\alpha_{1,2}$ represent the couplings of $Y_{1,2}$ particle to $Z^{\pm}_c \pi^\mp$ channel. The phases $\phi_{Z_{1,2}}$ fix the relative phases of the $Y_{1,2}$ resonances in the $Z^{\pm}_c$ chain contributions with respect to the scalar $\pi\pi/K\bar K$ amplitude.
The remaining $Y$ couplings are multiplicatively absorbed in $LS$ couplings. Furthermore, here we assumed $Z^{\pm}_c$ production undergoing strictly through $Y$ resonances. 

The interaction between the two light mesons has been shown to have a sizable impact on the invariant-mass distributions in the $e^+e^- \to J/\psi\pi^+\pi^-$ process \cite{BESIII:2017bua, Danilkin:2020kce}. Therefore, a proper treatment of the $\pi\pi/K\bar K$ final-state interaction (FSI) is required. Following our previous work~\cite{Ermolina:2024uln}, we describe the scalar-isoscalar FSI using the coupled-channel Omn\`es matrix. For the nonresonant production term we introduce
\begin{align}\label{eq:PiB}
\Pi_B^{\pi\pi}&=(a_B+b_B\sigma_2)\,\Omega^{(0)}_{11}(\sigma_2)+c_B\,\Omega^{(0)}_{12}(\sigma_2),\\
\Pi_B^{K\bar K}&=\frac{\sqrt{3}}{2}\left[(a_B+b_B\sigma_2)\,\Omega^{(0)}_{21}(\sigma_2)+c_B\,\Omega^{(0)}_{22}(\sigma_2)\right].\nonumber
\end{align}
Here $a_B$, $b_B$, and $c_B$ are real fit parameters. The additional factor $\sqrt{3}/2$ in the $K\bar K$ channel follows from the isospin normalization. The coupled-channel isospin $I=0$ Omn\`es matrix used in this work is based on the data-driven $N/D$ analysis of \cite{Danilkin:2020pak}, constrained by the Roy and Roy-Steiner analyses of $\pi\pi\to\pi\pi$ and $\pi\pi\to K\bar K$ scattering, respectively \cite{Garcia-Martin:2011iqs,Pelaez:2020gnd}. The scalar amplitudes associated with resonant production through a vector state $Y_{i}$ are parametrized as
\begin{align}
\Pi_{Y_{i}}^{\pi\pi}&=P_{Y_{i}}(s)\left[(a_{Y_{i}}+b_{Y_{i}}\sigma_2)\,\Omega^{(0)}_{11}(\sigma_2)+c_{Y_{i}}\,\Omega^{(0)}_{12}(\sigma_2)\right], \nonumber\\
\Pi_{Y_{i}}^{K\bar K}&=\frac{\sqrt{3}}{2}P_{Y_{i}}(s)\left[(a_{Y_{i}}+b_{Y_{i}}\sigma_2)\,\Omega^{(0)}_{21}(\sigma_2)+c_{Y_{i}}\,\Omega^{(0)}_{22}(\sigma_2)\right].\label{eq:PiR}
\end{align}
The full scalar contribution to the $e^+e^- \to J/\psi\pi^+\pi^-$ channel is then
\begin{equation}
\Pi^{\pi\pi}=e^{i\phi_B}\Pi_B^{\pi\pi}(\sigma_2)+\Pi_{Y_1}^{\pi\pi}(\sigma_2,s)+e^{i\phi_2}\Pi_{Y_2}^{\pi\pi}(\sigma_2,s),
\label{eq:Pi_total_pipi}
\end{equation}
where $\phi_B$ is the phase of the nonresonant background relative to the $Y(4220)$ contribution, and $\phi_2$ is the relative phase of the $Y(4320)$ contribution. The corresponding scalar contribution in the $e^+e^- \to J/\psi K^+K^-$ channel is
\begin{equation}
\Pi^{K\bar K}= e^{i\phi_B}\Pi_B^{K\bar K}(\sigma_2) + \Pi_{Y_1}^{K\bar K}(\sigma_2,s).
\label{eq:Pi_total_KK}
\end{equation}
We have explicitly verified that the contribution of the $Y(4320)$ term in the $J/\psi K^+K^-$ channel is negligible. The $\Pi^{\pi\pi}$ and $\Pi^{K\bar K}$ replace
\begin{align}\nonumber
    \alpha^{0 \to (31),2}_{01} X_0  \, \alpha^{(31) \to 3,1}_{00} \big|_{f_0(500)} + \alpha^{0 \to (31),2}_{01} X_0 \, \alpha^{(31) \to 3,1}_{00} \big|_{f_0(980)}
\end{align}
in the corresponding DPD expansions for $e^+e^- \to J/\psi\pi^+\pi^-$ and $e^+e^- \to J/\psi K^+K^-$, respectively.

Finally, the tensor $f_2(1270)$ contribution is parametrized as
\begin{align}\label{eq:f2_contribution}
&\alpha^{0 \to (31),2}_{01} X_2\,\alpha^{(31) \to 3,1}_{20} \\
&=P_{f_2(1270)}(\sigma_2)\left[f_1e^{i\phi_{f_1}}P_{Y_1}(s) +f_2 e^{i\phi_{f_2}}P_{Y_2}(s)\right] \nonumber,
\end{align}
where $f_{1,2}$ represent the couplings of the $f_2(1270)J/\psi$ channel to the $Y_{1,2}$, and $\phi_{f_1}$, $\phi_{f_2}$ are the corresponding relative phases. We have also tested a nonresonant tensor background term and found it to be negligible within the present data precision; it is therefore not included in the nominal fit. 

We note that in the present work the $s$ dependence of the production amplitude is encoded through the explicit $Y(4220)$ and $Y(4320)$ propagators with constant widths $\Gamma_R$, together with a smooth nonresonant background term. Possible open-charm loop mechanisms, including triangle singularities, are not incorporated explicitly. An analysis of their significance is provided by Ref.~\cite{Chen:2026fnz}, which results in decline of $\chi^2/N_{\text{d.o.f.}}=1.43$--$1.57$ due to the removal of triangle diagrams (d.o.f. defines the degrees of freedom). We therefore do not expect such mechanisms to be the dominant source of structures in the region $4.13 \le \sqrt{s} \,(\mathrm{GeV}) \le 4.36$, and assume their residual effect to be partly absorbed into the effective production amplitudes introduced above; we return to this point in Sec.~\ref{sec:comparison}. A dedicated treatment of such mechanisms is left for future work.

In the next section, we proceed with the application of the defined model to describe the $e^+ e^- \to J/\psi \, \pi \, \pi \, (K \bar{K})$ data.

\section{Results and discussion}
\label{sec3}

Before comparing the model to the one-dimensional invariant-mass distributions, the experimental spectra are converted to differential cross sections. For each center-of-mass energy $\sqrt{s}$ and each invariant-mass variable $m_{ij}$, the corresponding histogram is normalized such that its integrated value reproduces the measured Born cross section $\sigma_{\rm exp}(\sqrt{s})$ for the full process.

\subsection{Minimal fit}

\begin{figure*}[htbp!]
	\centering 
	\includegraphics[width=0.47\textwidth]{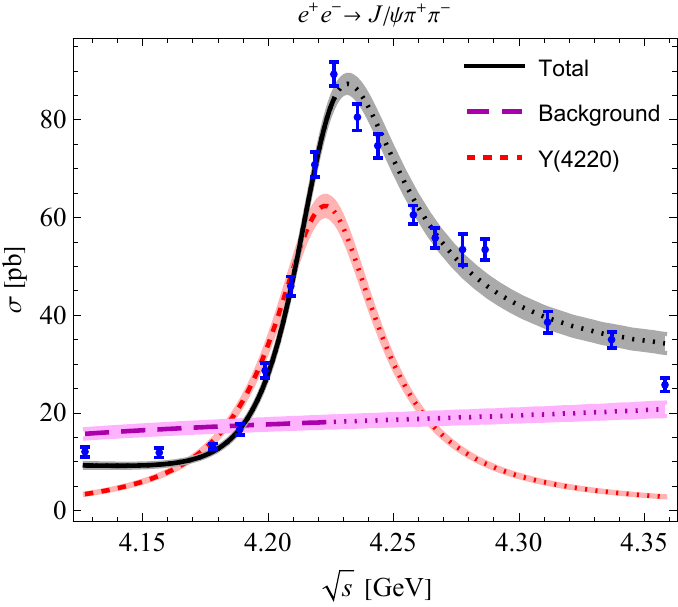} 
	\includegraphics[width=0.47\textwidth]{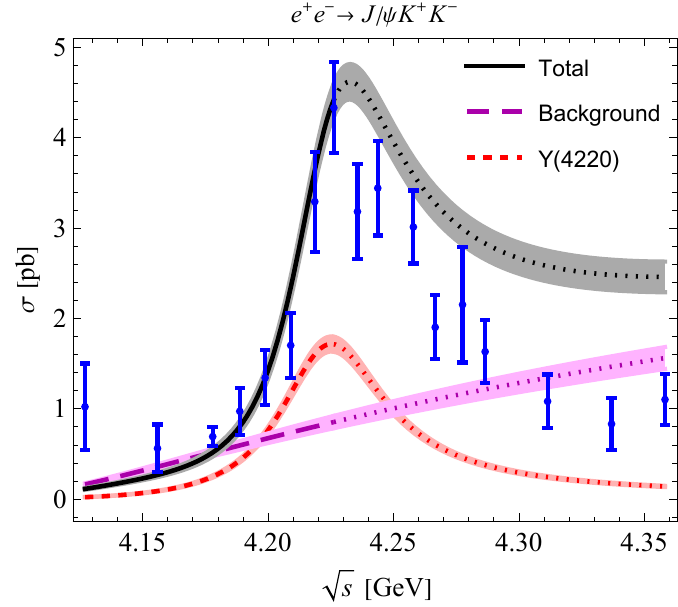}
	\caption{Total cross section of the process $e^+ e^- \to J/\psi \, \pi \, \pi \, (K \bar{K})$ (solid, black), the $Y(4220)$ resonant contribution (red, dashed), and nonresonant background contribution (purple, long dashed) in the \textit{minimal fit}. The data are taken from \cite{BESIII:2022qal, BESIII:2022joj}. The predictions for each contribution are shown as dotted lines. The shaded bands correspond to $2\sigma$ uncertainties.}
	\label{fig_1}
\end{figure*}

\begin{figure*}[htbp!]
	\centering 
	\includegraphics[width=0.93\textwidth]{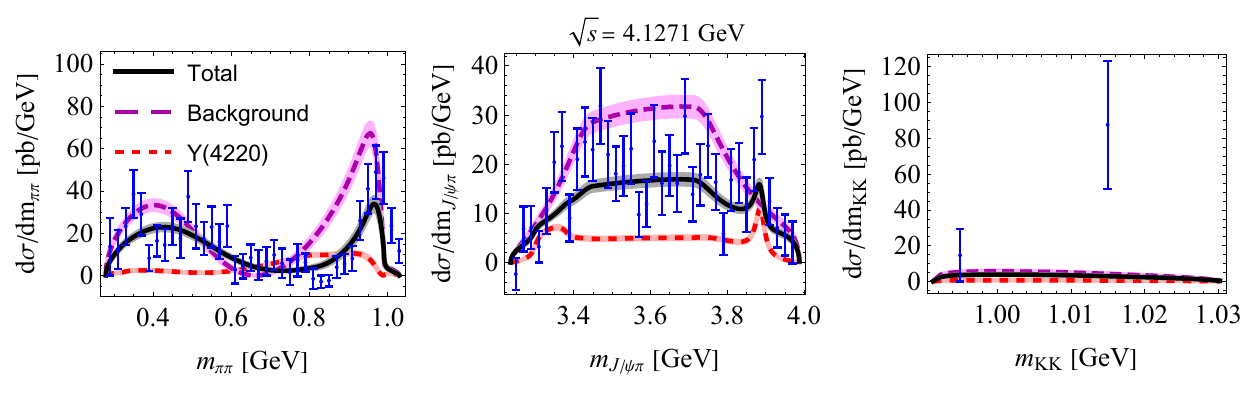}\\
    \includegraphics[width=0.93\textwidth]{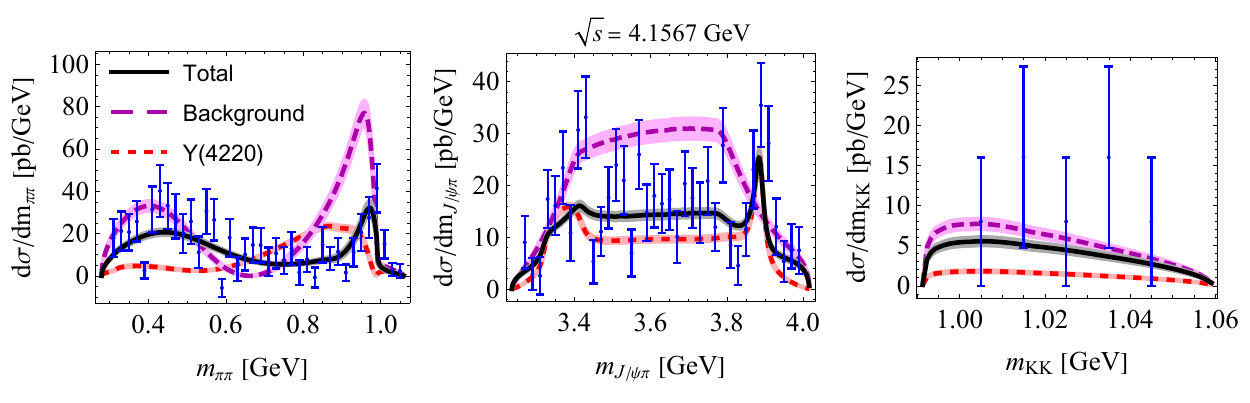}\\
    \includegraphics[width=0.93\textwidth]{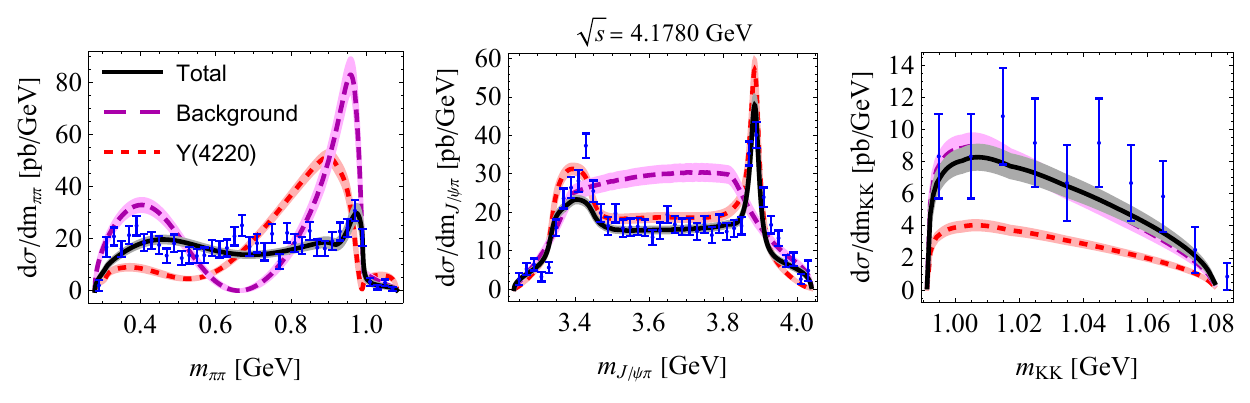}\\
    \includegraphics[width=0.93\textwidth]{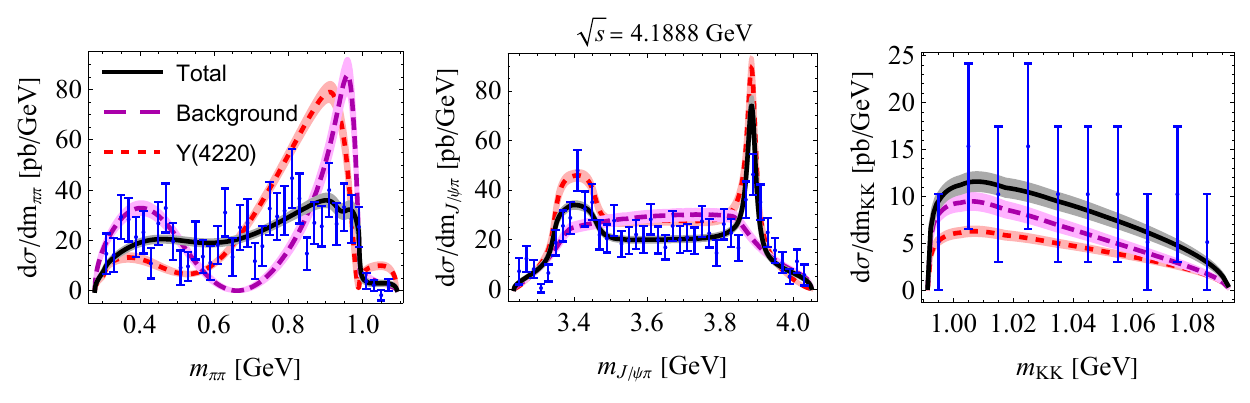}
    \caption{\textit{Minimal fit} to the invariant-mass distributions of the $e^+ e^- \to J/\psi \, \pi^+ \, \pi^- \, (K\bar{K})$ process at the c.m. energies $\sqrt{s}=4.1271$--$4.1888$ GeV. The data are taken from \cite{BESIII:2025qkn, BESIII:2022joj}. In all panels, the solid black curve is the full result, the long dashed purple curve is the nonresonant background, and the dashed red curve is the $Y(4220)$ contribution. The shaded bands correspond to $2\sigma$ uncertainties.}
    \label{fig_3.1}
\end{figure*}

\begin{figure*}[htbp!]
	\centering 
    \includegraphics[width=0.93\textwidth]{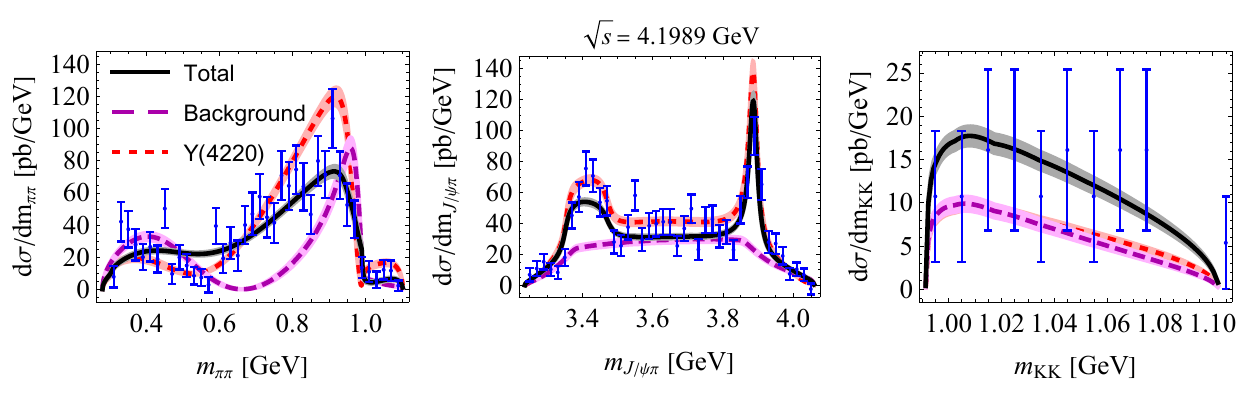}\\
    \includegraphics[width=0.93\textwidth]{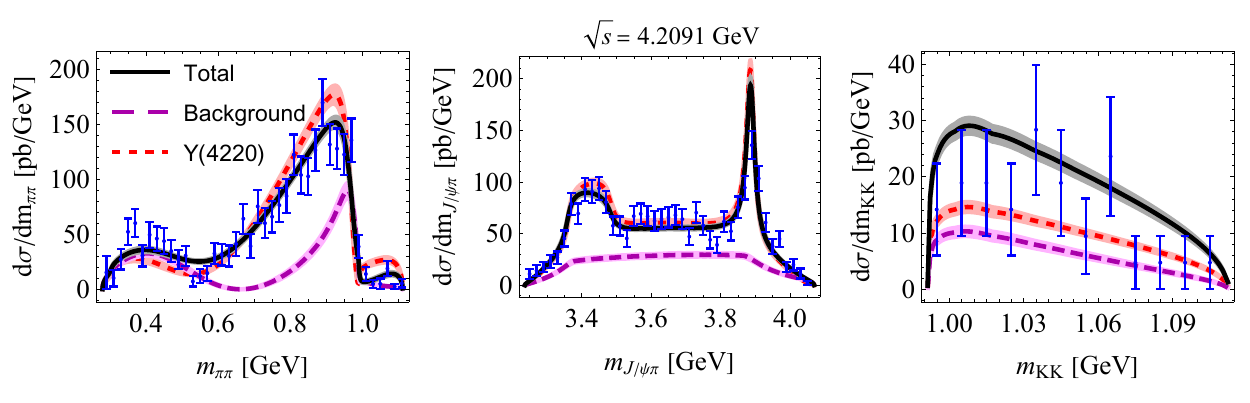}\\
    \includegraphics[width=0.93\textwidth]{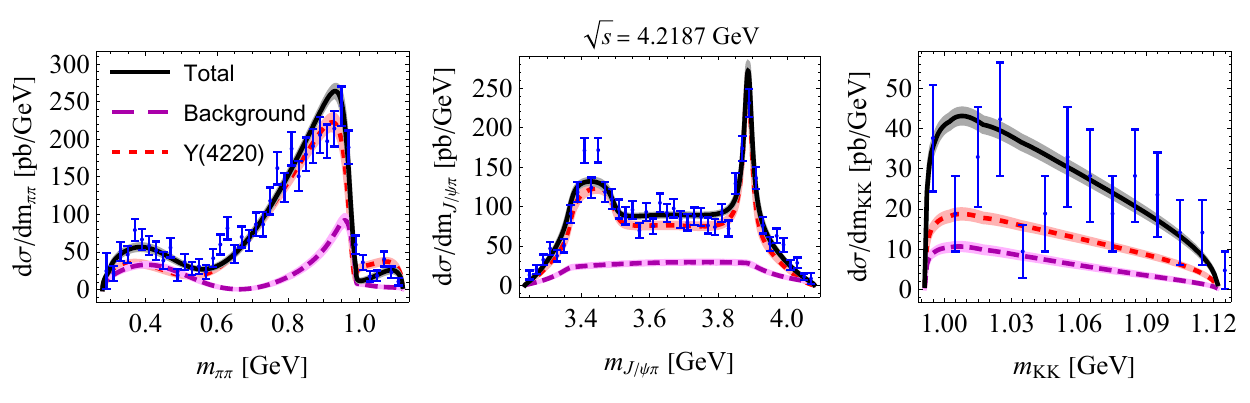}\\
    \includegraphics[width=0.93\textwidth]{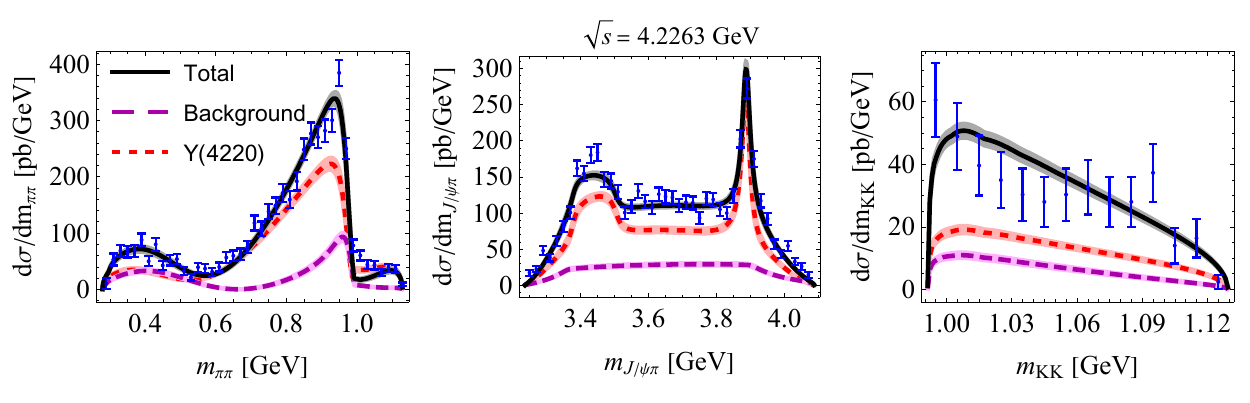}
	\caption{\textit{Minimal fit} to the invariant-mass distributions of the $e^+ e^- \to J/\psi \, \pi^+ \, \pi^- \, (K\bar{K})$ process at the c.m. energies $\sqrt{s}=4.1989$--$4.2263$ GeV. The data are taken from \cite{BESIII:2025qkn, BESIII:2022joj}. In all panels, the solid black curve is the full result, the long dashed purple curve is the nonresonant background, and the dashed red curve is the $Y(4220)$ contribution. The shaded bands correspond to $2\sigma$ uncertainties.}
	\label{fig_3.2}
\end{figure*}

We first consider a restricted dataset with $4.1271 \le \sqrt{s} \,(\mathrm{GeV}) \le 4.2263$, corresponding to the energy region dominated by the first vector structure. In this minimal fit we retain only the $Y(4220)$ resonant production, the $Z_c^\pm(3900)$ contribution in the $J/\psi\pi^\pm$ subsystems, and the scalar-isoscalar $\pi\pi/K\bar K$ $S$-wave described by the coupled-channel Omn\`es representation. The $Y(4320)$ and tensor $f_2(1270)$ contributions are omitted. The fit contains nine parameters, listed in Table~\ref{tab2}. To account for an underestimation of total cross section uncertainties arising from invariant-mass distributions compared to the uncertainties of total cross section measurements, hereafter we provide a $2\sigma$ error (unless otherwise stated), with this conservative scaling aligning the parameter uncertainties with the larger total cross section uncertainties. The comparison of cross section values at the most statistically significant energy point is provided in Table~\ref{tab22}. The masses and widths of the $Y(4220)$ and $Z_c(3900)$ are fixed to the PDG values. The fit gives $\chi^2/N_{\rm d.o.f.}=1.8$. The corresponding total cross sections are shown in Fig.~\ref{fig_1}, and the one-dimensional invariant-mass distributions are compared with the data in Figs.~\ref{fig_3.1} and \ref{fig_3.2}.

\begin{table}[h]
\centering
\renewcommand*{\arraystretch}{1.4}
\begin{tabular}{l r@{\;$\pm$\;}l | l r@{\;$\pm$\;}l}
 \hline 
 Parameter & \multicolumn{2}{c|}{\phantom{0}Value} & Parameter & \multicolumn{2}{c}{\phantom{0}Value} \\ 
 \hline 
 $\alpha_1 \ (\text{GeV}^2)$ & $1.62$ & $0.05$ & $\phi_{Z_1} \ (\text{rad})$ & $0.6$ & $0.4$ \\ 
 $a_B$ & $-36.9$ & $1.3$ & $a_{Y_1}$ & $9.1$ & $0.5$ \\ 
 $b_B \ (\text{GeV}^{-2})$ & $60.6$ & $3.0$ & $b_{Y_1} \ (\text{GeV}^{-2})$ & $-24.3$ & $0.8$ \\ 
 $c_B$ & $29.8$ & $1.2$ & $c_{Y_1}$ & $-10.8$ & $0.3$ \\ 
 $\phi_B \ (\text{rad})$ & $3.10$ & $0.04$ & \multicolumn{3}{c}{} \\ 
 \hline
 \multicolumn{6}{c}{$\chi^2/N_{\text{d.o.f.}} = 1.8$} \\
 \hline
\end{tabular}
\caption{Fit parameters of the \textit{minimal fit} with $2\sigma$ errors indicating the uncertainties in the experimental data.}
\label{tab2}
\end{table}

\begin{table}[h]
\centering
\begin{tabular}{lcc}
\toprule
Process & Source & Cross section (pb) \\
\midrule
\multirow{3}{*}{$e^+e^- \to J/\psi\pi^+\pi^-$}
 & BESIII total & $89.29 \pm 2.46$ \\   
 & $m_{\pi\pi}$ & $89.29 \pm 1.62$ \\
 & $m_{J/\psi}$ & $89.29 \pm 1.14$ \\
\midrule
\multirow{2}{*}{$e^+e^- \to J/\psi K^+K^-$}
 & BESIII total & $4.33 \pm 0.50$ \\
 & $m_{KK}$ & $4.33 \pm 0.32$ \\
\bottomrule
\end{tabular}
\caption{Comparison of $1\sigma$ cross section uncertainty from the normalized invariant-mass distributions \cite{BESIII:2025qkn, BESIII:2022joj} and BESIII analyses of the total cross sections \cite{BESIII:2022qal, BESIII:2022joj} for the energy $\sqrt{s}=4.2263$ GeV.}
\label{tab22}
\end{table}

\begin{figure*}[htbp!]
	\centering 
    \includegraphics[width=0.93\textwidth]{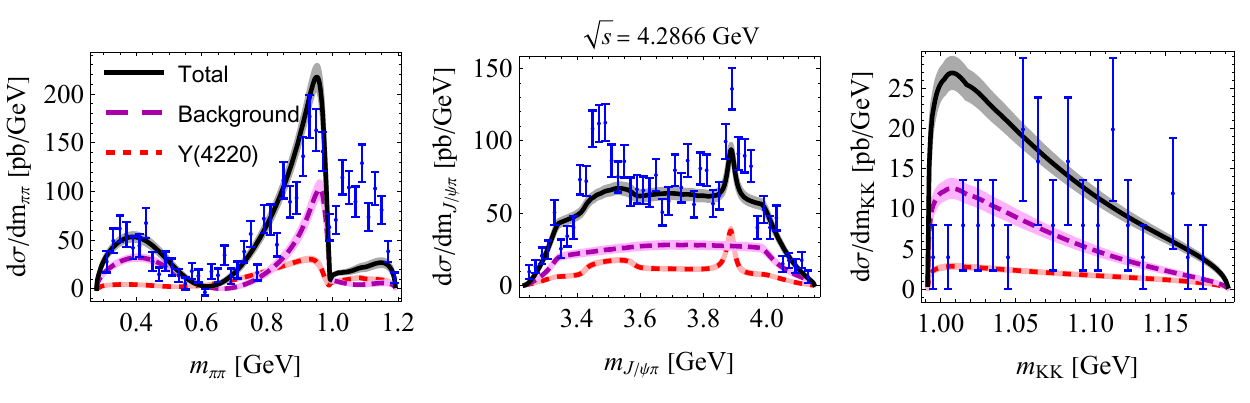}
	\caption{\textit{Minimal-fit extrapolation} to the invariant-mass distributions of the $e^+ e^- \to J/\psi \, \pi^+ \, \pi^- \, (K\bar{K})$ process at the c.m. energy $\sqrt{s}=4.2866$ GeV. The data are taken from \cite{BESIII:2025qkn, BESIII:2022joj}. The solid black curve is the full result, the long dashed purple curve is the nonresonant background, and the dashed red curve is the $Y(4220)$ contribution. The shaded bands correspond to $2\sigma$ uncertainties.}
	\label{fig_33}
\end{figure*}

The comparison shows that the nonresonant scalar production term, followed by $\pi\pi/K\bar K$ rescattering, is essential for describing the invariant-mass distributions. This is particularly visible at the lower energies, where the nonresonant scalar component gives a sizable contribution to the $m_{\pi\pi}$ and $m_{K\bar K}$ spectra. The restricted model also provides a useful extrapolation up to 
the data point corresponding to 
$\sqrt{s}=4.2866~\mathrm{GeV}$, as shown in Fig.~\ref{fig_33}. At higher energies, however, the additional $Y(4320)$ and $f_2(1270)$ contributions become necessary, motivating the total fit described in the next subsection.

This restricted fit is similar in spirit to Ref.~\cite{vonDetten:2024eie}, where the data in the $4.2$-$4.35~\mathrm{GeV}$ region are described without introducing an additional exotic vector state near $4.32~\mathrm{GeV}$, with the approach further developed in Ref.~\cite{Dong:2026fdi}. In our case, the open-charm meson-loop mechanisms of Refs.~\cite{vonDetten:2024eie, Dong:2026fdi} are not included explicitly, but are effectively absorbed into the nonresonant scalar production term. In contrast, these analyses retaining the open-charm loop mechanisms simplify the light-meson final-state interaction. More details on the model comparison can be found in Sec.~\ref{sec:comparison}.


\subsection{Total fit}

\begin{table}[t!]
\centering
\renewcommand*{\arraystretch}{1.4}
\begin{tabular}{l r@{\;$\pm$\;}l | l r@{\;$\pm$\;}l}
 \hline 
 Parameter & \multicolumn{2}{c|}{\phantom{0}Value} & Parameter & \multicolumn{2}{c}{\phantom{0}Value} \\ 
 \hline 
 $\alpha_1 \ (\text{GeV}^2)$ & $1.11$ & $0.13$ & $\alpha_2 \ (\text{GeV}^2)$ & $2.1$ & $0.3$ \\ 
 $a_B$ & $24.3$ & $1.4$ & $a_{Y_2}$ & $-1.4$ & $1.4$ \\ 
 $b_B \ (\text{GeV}^{-2})$ & $-21.2$ & $4.1$ & $b_{Y_2} \ (\text{GeV}^{-2})$ & $11.4$ & $3.4$ \\ 
 $c_B$ & $-11.9$ & $1.4$ & $c_{Y_2}$ & $6.2$ & $1.1$ \\ 
 $\phi_B \ (\text{rad})$ & $2.56$ & $0.07$ & $\phi_{Z_2} \ (\text{rad})$ & $0.8$ & $0.2$ \\ 
 $\phi_{Z_1} \ (\text{rad})$ & $2.6$ & $0.3$ & $\phi_{2} \ (\text{rad})$ & $2.1$ & $0.2$ \\ 
 $a_{Y_1}$ & $-5.9$ & $0.4$ & $f_2 \ (\text{GeV}^2)$ & $-25.9$ & $2.7$ \\ 
 $b_{Y_1} \ (\text{GeV}^{-2})$ & $16.3$ & $1.1$ & $\phi_{f_2} \ (\text{rad})$ & $1.1$ & $0.3$ \\ 
 $c_{Y_1}$ & $7.8$ & $0.5$ & $\phi_{f_1} \ (\text{rad})$ & $1.1$ & $0.3$ \\ 
 $f_1 \ (\text{GeV}^2)$ & $9.8$ & $1.8$ & & \multicolumn{2}{c}{} \\ 
 \hline
 \multicolumn{6}{c}{$\chi^2/N_{\text{d.o.f.}} = 1.66$} \\
 \hline
\end{tabular}
\caption{Fit parameters of the \textit{total fit} with $2\sigma$ errors indicating the uncertainties in the experimental data.}
\label{tab3}
\end{table}

We now fit the full dataset in the range $4.1271 \le \sqrt{s} \,(\mathrm{GeV}) \le 4.3583$. In addition to the ingredients of the minimal fit, we include the second vector structure $Y(4320)$ in the $J/\psi\pi^+\pi^-$ channel and the tensor $f_2(1270)$ contribution in the $\pi^+\pi^-$ subsystem. The fit contains 19 parameters, describing the resonant and nonresonant production mechanisms, summarized in Table~\ref{tab3}. In addition, the masses and widths of the $Z_c(3900)$, $Y(4220)$, and $Y(4320)$ are allowed to vary, giving 25 fit parameters in total. The scalar-isoscalar $\pi\pi/K\bar K$ final-state interaction is kept fixed by the Omn\`es input described in Sec.~\ref{sec2}. The current data do not allow us to constrain the $Z_c(4020)$ contribution reliably; it is therefore not included in the nominal fit, although the formalism can accommodate it once more precise high-energy data become available.

\begin{figure*}[t]
	\centering 
	\includegraphics[width=0.47\textwidth]{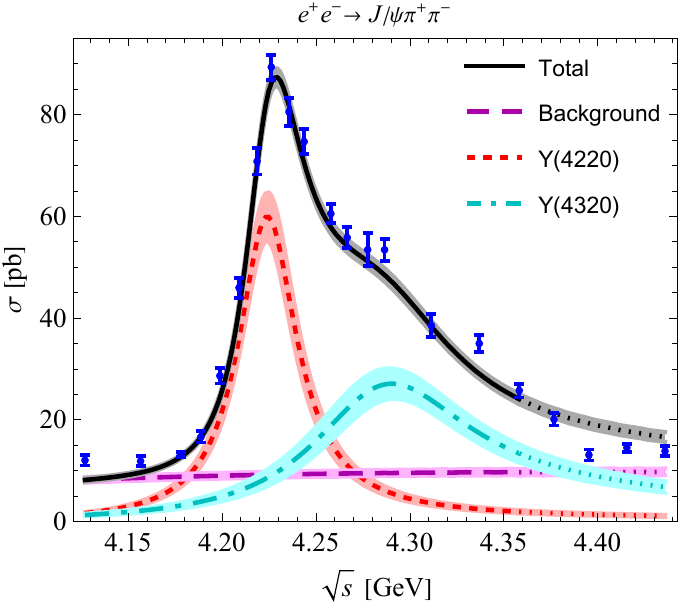} 
	\includegraphics[width=0.47\textwidth]{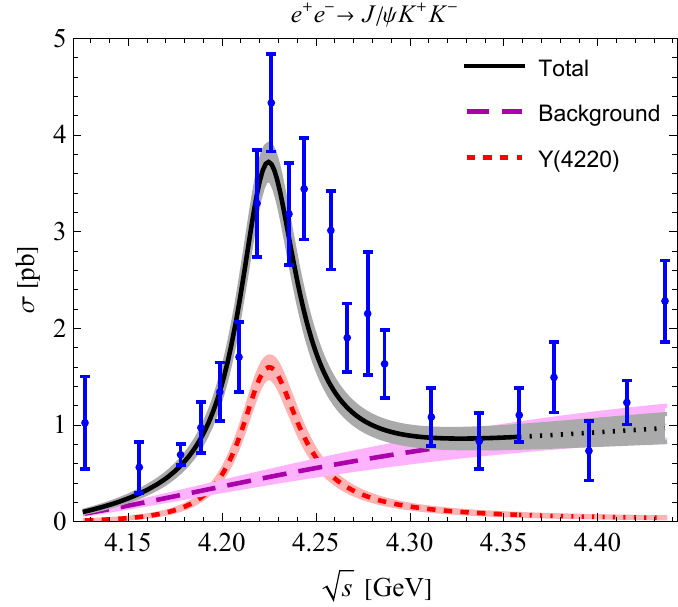}
    \caption{Total cross section of the process $e^+ e^- \to J/\psi \, \pi \, \pi \, (K \bar{K})$ (solid, black), the $Y(4220)$ resonant contribution (red, dashed), the $Y(4320)$ resonant contribution (cyan, dot-dashed) and nonresonant background contribution (purple, long dashed) in the \textit{total fit}. The data are taken from \cite{BESIII:2022qal, BESIII:2022joj}. The predictions for each contribution are shown as dotted lines. The shaded bands correspond to $2\sigma$ uncertainties.}
	\label{fig:total_cross_sections}
\end{figure*}

\begin{figure*}[t]
	\centering 
	\includegraphics[width=0.47\textwidth]{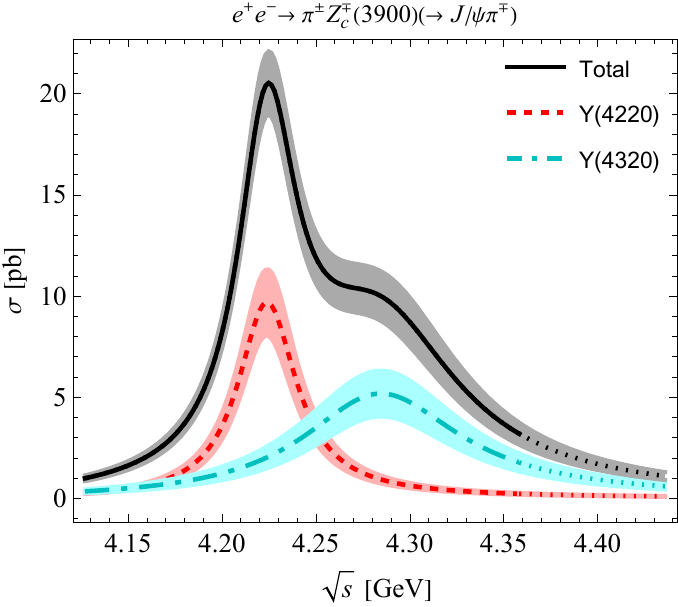}
	\includegraphics[width=0.47\textwidth]{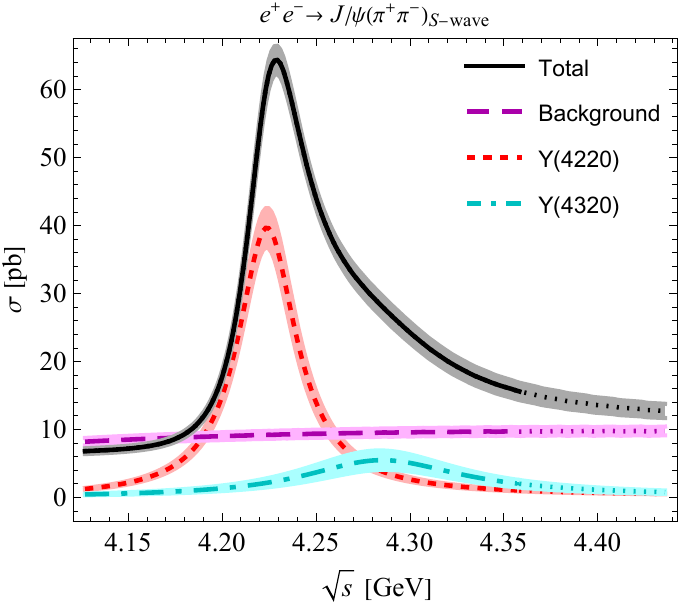}
    \caption{Total cross section of the subprocesses of $e^+ e^- \to J/\psi \, \pi \, \pi \, (K \bar{K})$ (solid, black), the $Y(4220)$ resonant contribution (red, dashed), the $Y(4320)$ resonant contribution (cyan, dot-dashed) and nonresonant background contribution (purple, long dashed) in the \textit{total fit}. The predictions for each contribution are shown as dotted lines. The shaded bands correspond to $2\sigma$ uncertainties.}
	\label{fig:subprocess_cross_sections}
\end{figure*}

\begin{figure*}[htbp!]
	\centering 
	\includegraphics[width=0.93\textwidth]{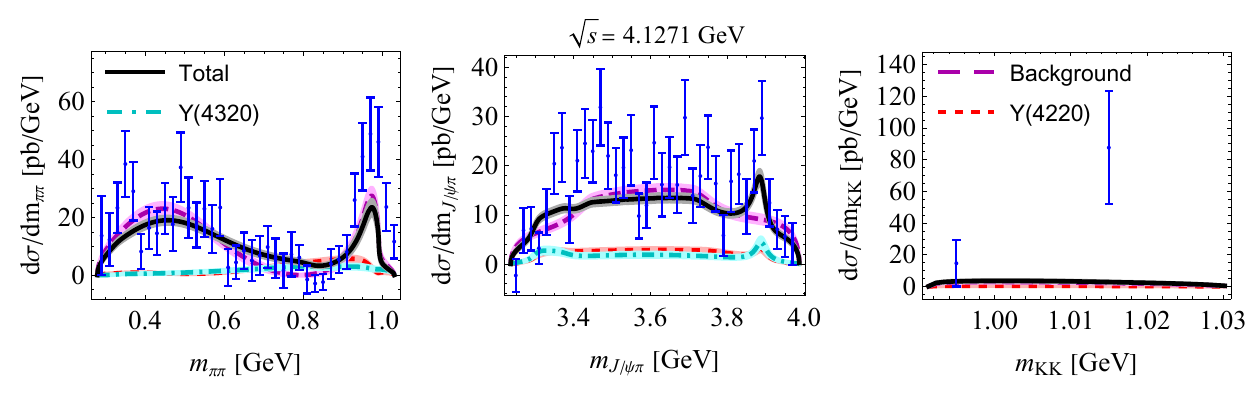}\\
    \includegraphics[width=0.93\textwidth]{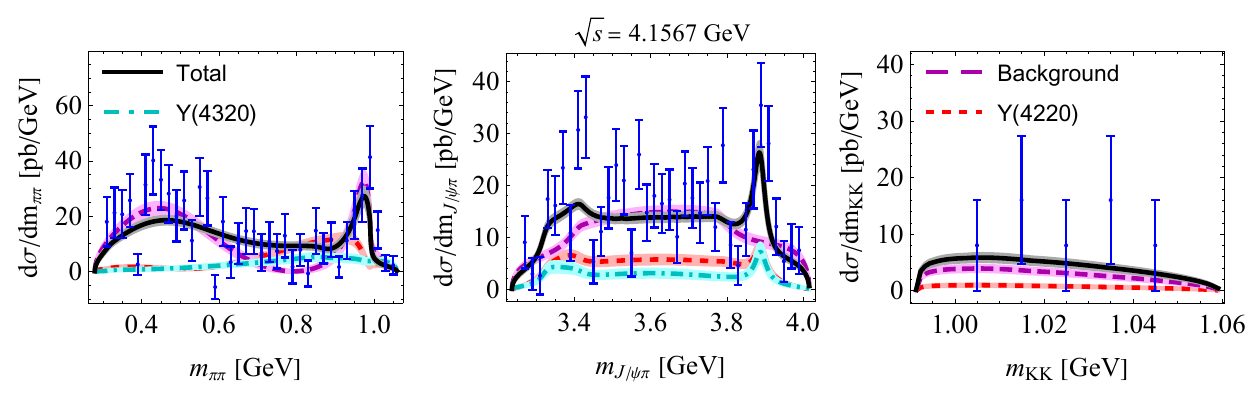}\\
    \includegraphics[width=0.93\textwidth]{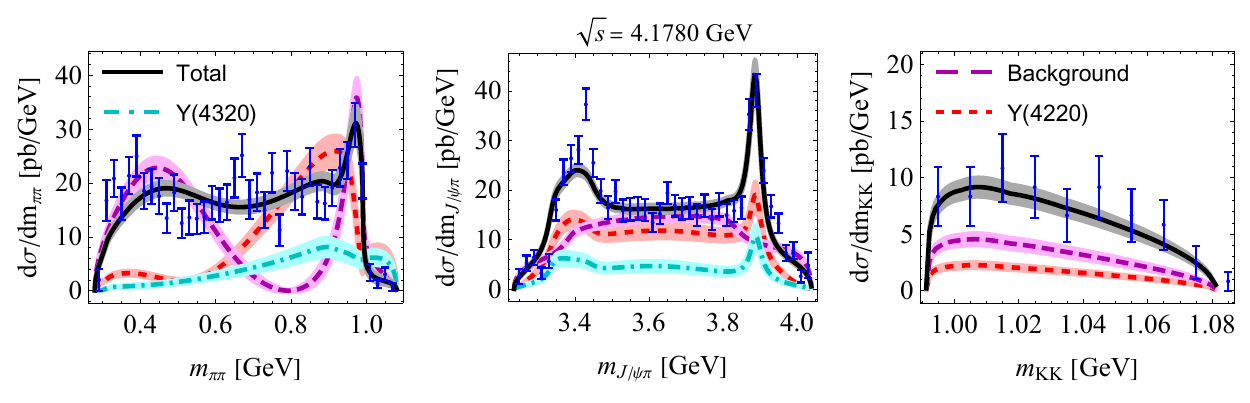}\\
    \includegraphics[width=0.93\textwidth]{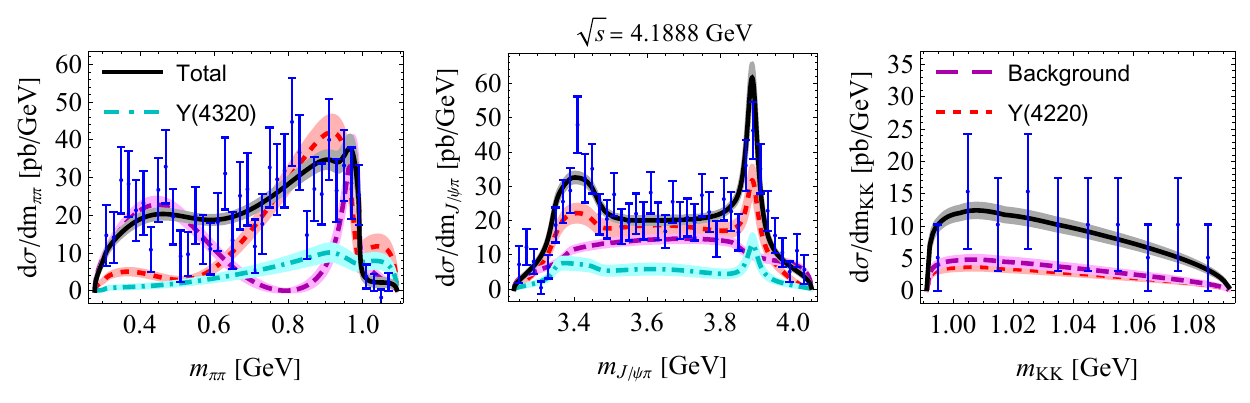}
    \caption{\textit{Total fit} to the invariant-mass distributions of the $e^+ e^- \to J/\psi \, \pi^+ \, \pi^- \, (K\bar{K})$ process at the c.m. energies $\sqrt{s}=4.1271$--$4.1888$ GeV. The data are taken from \cite{BESIII:2025qkn, BESIII:2022joj}. In all panels, the solid black curve is the full result, the long dashed purple curve is the nonresonant background, the dashed red curve is the $Y(4220)$ contribution, and the dot-dashed cyan curve is the $Y(4320)$ contribution when present. The shaded bands correspond to $2\sigma$ uncertainties.}
    \label{fig_6.1}
\end{figure*}

\begin{figure*}[htbp!]
	\centering 
    \includegraphics[width=0.93\textwidth]{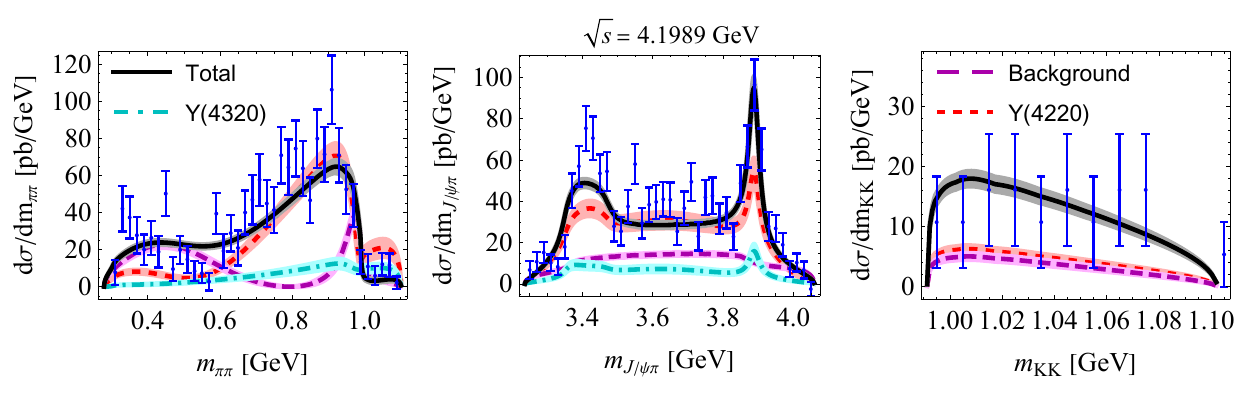}\\
    \includegraphics[width=0.93\textwidth]{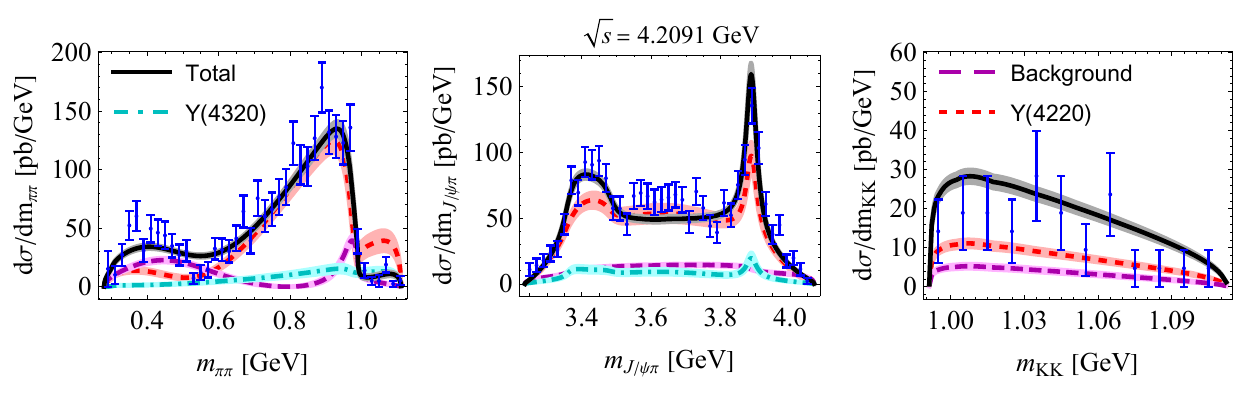}\\
    \includegraphics[width=0.93\textwidth]{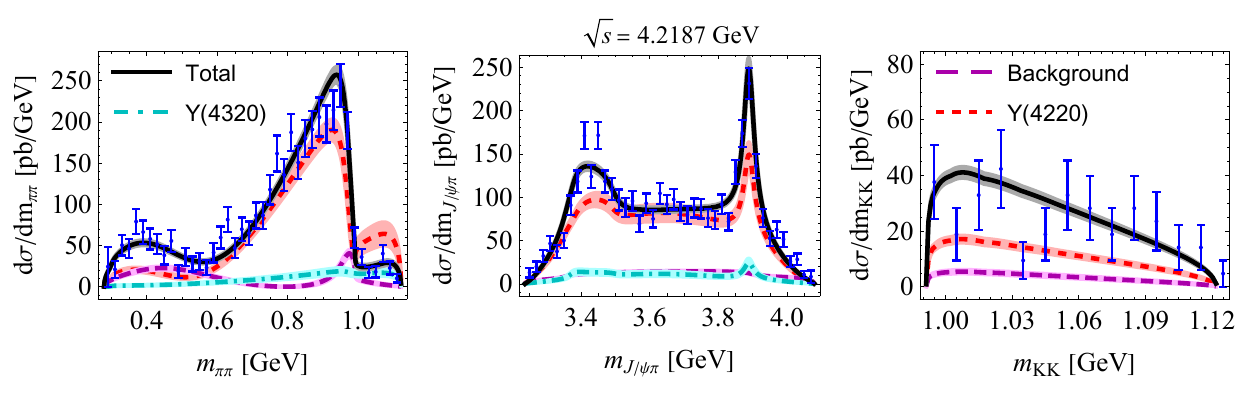}\\
    \includegraphics[width=0.93\textwidth]{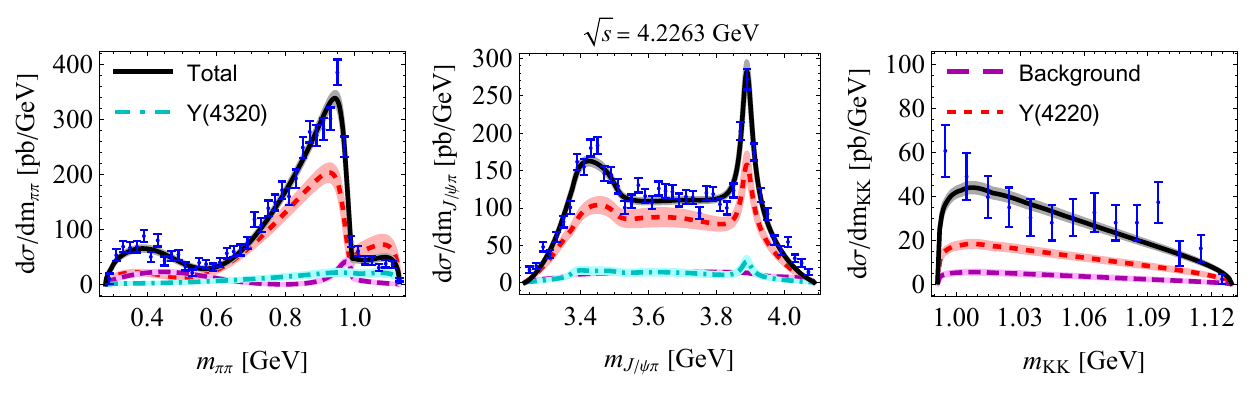}
    \caption{\textit{Total fit} to the invariant-mass distributions of the $e^+ e^- \to J/\psi \, \pi^+ \, \pi^- \, (K\bar{K})$ process at the c.m. energies $\sqrt{s}=4.1989$--$4.2263$ GeV. The data are taken from \cite{BESIII:2025qkn, BESIII:2022joj}. In all panels, the solid black curve is the full result, the long dashed purple curve is the nonresonant background, the dashed red curve is the $Y(4220)$ contribution, and the dot-dashed cyan curve is the $Y(4320)$ contribution when present. The shaded bands correspond to $2\sigma$ uncertainties.}
    \label{fig_6.2}
\end{figure*}

\begin{figure*}[htbp!]
	\centering 
    \includegraphics[width=0.93\textwidth]{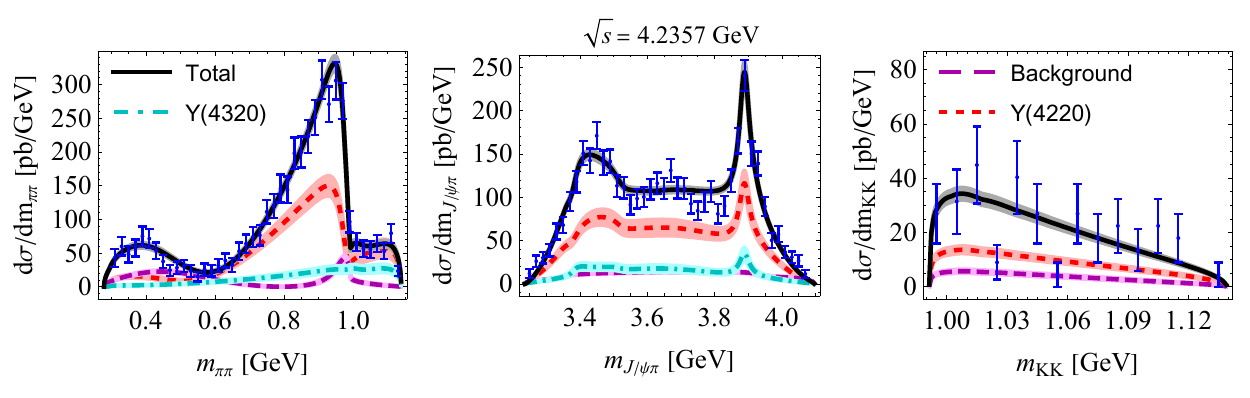}\\
    \includegraphics[width=0.93\textwidth]{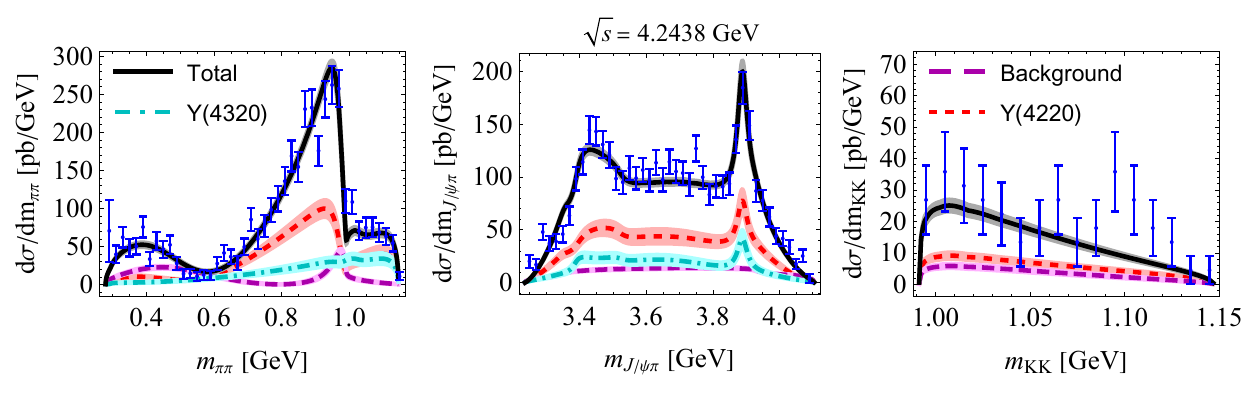}\\
    \includegraphics[width=0.93\textwidth]{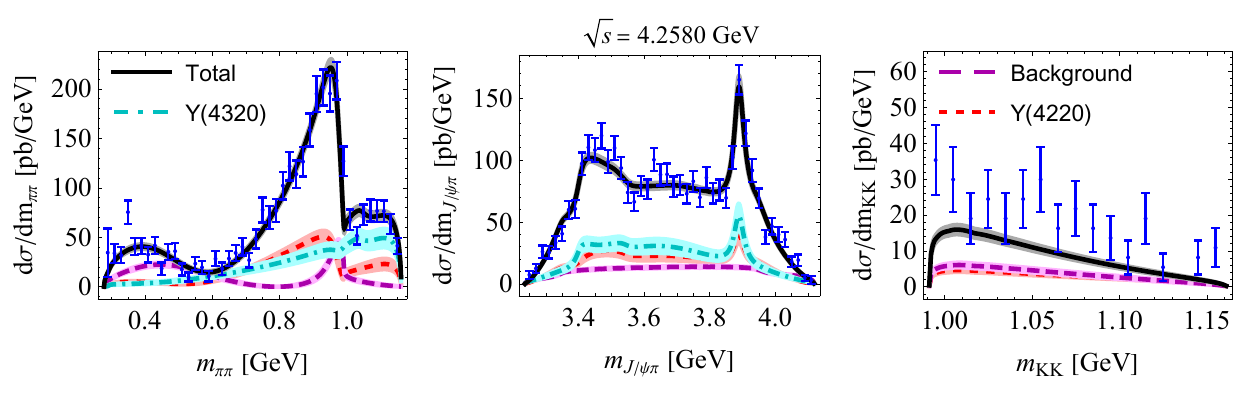}\\
    \includegraphics[width=0.93\textwidth]{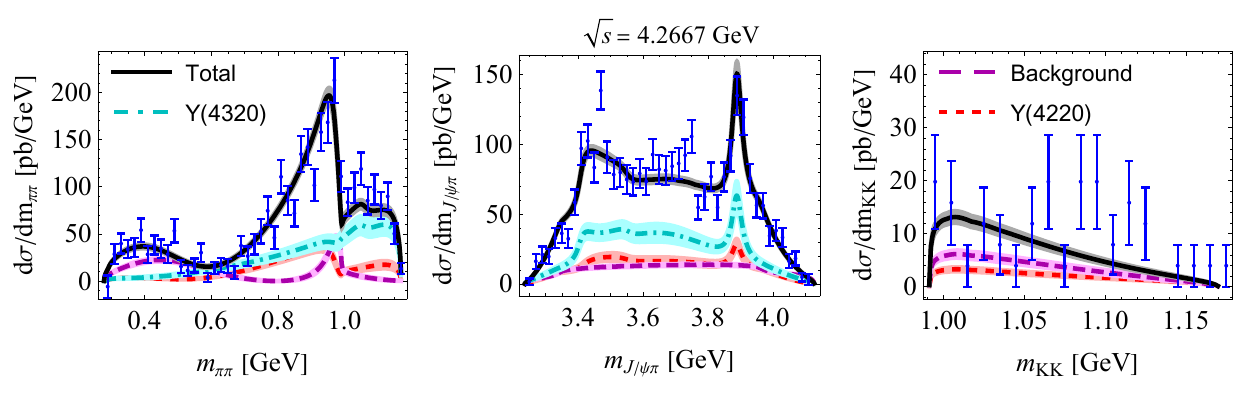}
    \caption{\textit{Total fit} to the invariant-mass distributions of the $e^+ e^- \to J/\psi \, \pi^+ \, \pi^- \, (K\bar{K})$ process at the c.m. energies $\sqrt{s}=4.2357$--$4.2667$ GeV. The data are taken from \cite{BESIII:2025qkn, BESIII:2022joj}. In all panels, the solid black curve is the full result, the long dashed purple curve is the nonresonant background, the dashed red curve is the $Y(4220)$ contribution, and the dot-dashed cyan curve is the $Y(4320)$ contribution when present. The shaded bands correspond to $2\sigma$ uncertainties.}
    \label{fig_6.3}
\end{figure*}

\begin{figure*}[htbp!]
	\centering 
    \includegraphics[width=0.93\textwidth]{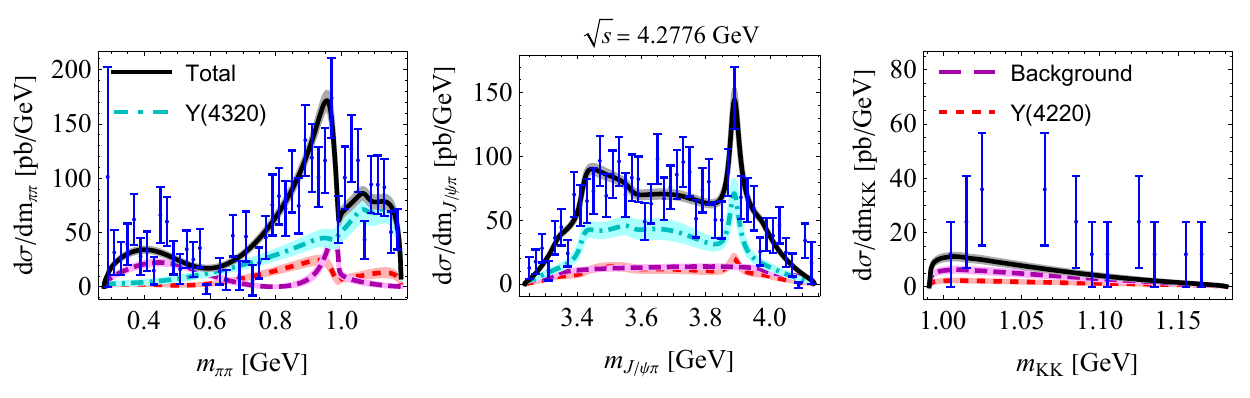}\\
    \includegraphics[width=0.93\textwidth]{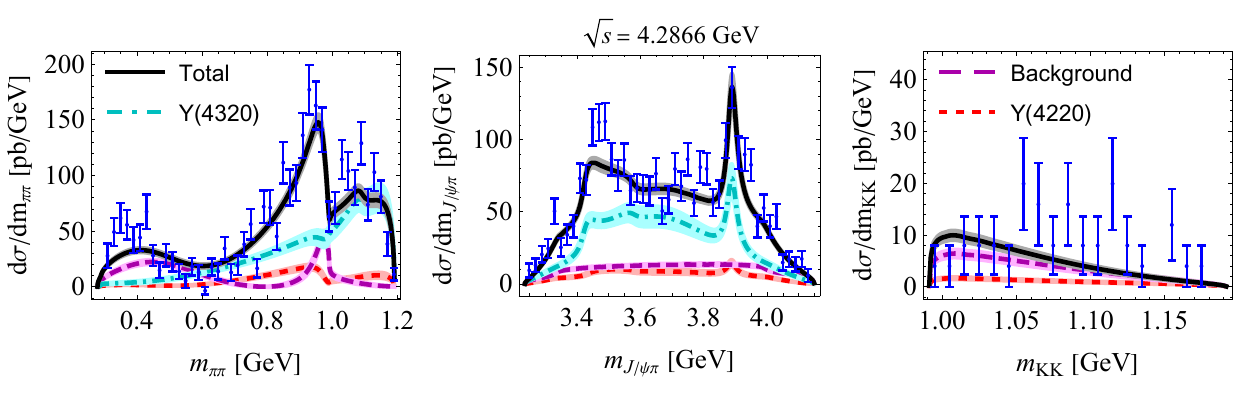}\\
    \includegraphics[width=0.93\textwidth]{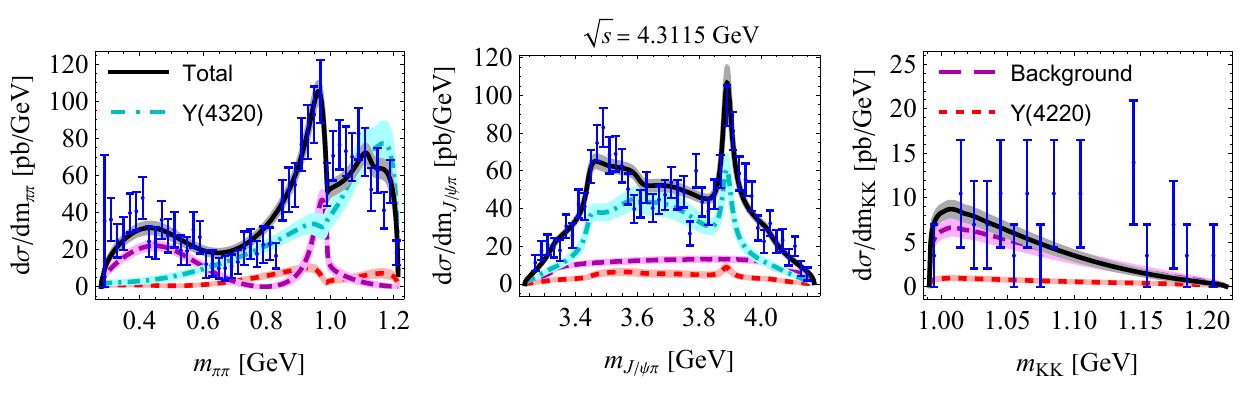}\\
    \includegraphics[width=0.93\textwidth]{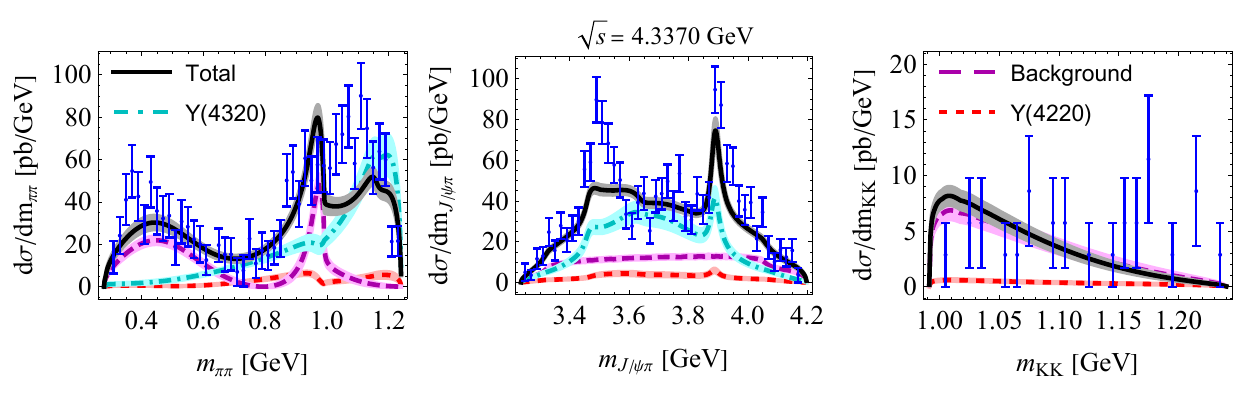}
	\caption{\textit{Total fit} to the invariant-mass distributions of the $e^+ e^- \to J/\psi \, \pi^+ \, \pi^- \, (K\bar{K})$ process at the c.m. energies $\sqrt{s}=4.2776$--$4.3370$ GeV. The data are taken from \cite{BESIII:2025qkn, BESIII:2022joj}. In all panels, the solid black curve is the full result, the long dashed purple curve is the nonresonant background, the dashed red curve is the $Y(4220)$ contribution, and the dot-dashed cyan curve is the $Y(4320)$ contribution when present. The shaded bands correspond to $2\sigma$ uncertainties.}
    \label{fig_6.4}
\end{figure*}

\begin{figure*}[htbp!]
	\centering
    \includegraphics[width=0.93\textwidth]{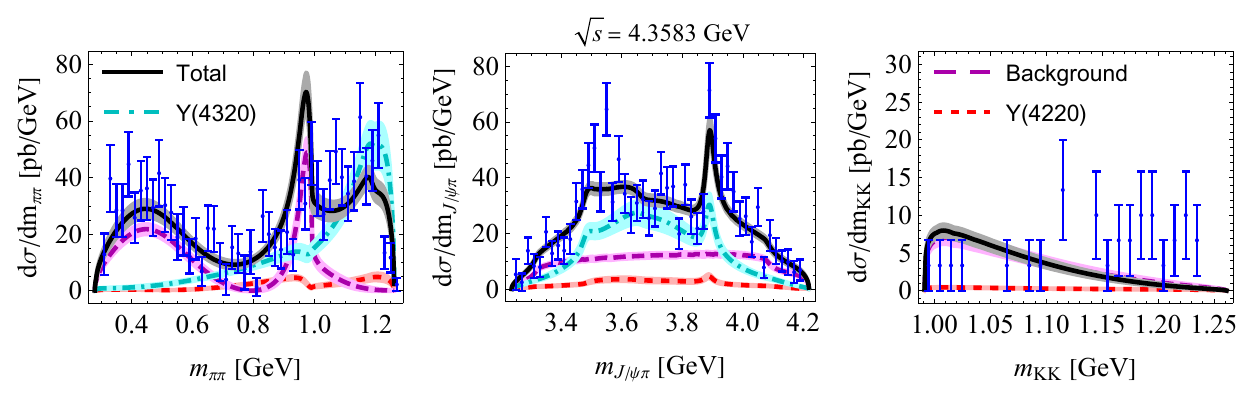}
	\caption{\textit{Total fit} to the invariant-mass distributions of the $e^+ e^- \to J/\psi \, \pi^+ \, \pi^- \, (K\bar{K})$ process at the c.m. energy $\sqrt{s}=4.3583$ GeV. The data are taken from \cite{BESIII:2025qkn, BESIII:2022joj}. In all panels, the solid black curve is the full result, the long dashed purple curve is the nonresonant background, the dashed red curve is the $Y(4220)$ contribution, and the dot-dashed cyan curve is the $Y(4320)$ contribution when present. The shaded bands correspond to $2\sigma$ uncertainties.}
    \label{fig_6.5}
\end{figure*}

The total fit gives $\chi^2/N_{\rm d.o.f.}=1.66$. The resulting total cross sections are shown in Fig.~\ref{fig:total_cross_sections}. The extracted subprocess cross sections for $e^+e^-\to \pi^\pm Z_c^\mp(3900)\to J/\psi\pi^+\pi^-$ and for the scalar $J/\psi(\pi^+\pi^-)_{S\text{-wave}}$ channel are shown in Fig.~\ref{fig:subprocess_cross_sections}. The invariant-mass distributions are shown in Figs.~\ref{fig_6.1}-\ref{fig_6.5} The sizable interference between the resonant and nonresonant scalar amplitudes demonstrates that a purely resonant description is not sufficient for the BESIII data. Compared with the minimal fit, the nonresonant scalar background is reduced once the $Y(4320)$ and $f_2(1270)$ contributions are included. \footnote{As seen from the $e^+ e^- \to J/\psi \, \pi \, \pi \, (K \bar{K})$ cross section, the first two data points contain only the tails of the resonant $Y$ states, which results in a small contribution from the exotic $Z_c$ state to the invariant-mass distributions. This indicates additional contributions from conventional charmonium states, the explicit modeling of which is beyond the scope of this work.}
If the model is to be extrapolated beyond the fitted energy range, a more detailed modeling of the $s$ dependence of the nonresonant production amplitude will be necessary.

\begin{table}[h!]
\centering
\begin{tabular}{llll}
\toprule
State & Source & Mass (MeV) & Width (MeV) \\
\midrule
\multirow{3}{*}{$Z_c^\pm(3900)$}
 & This work & $3888.7 \pm 1.8$ & $37.0 \pm 2.9$ \\
 & BESIII PWA
 & $3884.6 \pm 0.7 \pm 3.3$ & $37.2 \pm 1.3 \pm 6.6$ \\
 & PDG
 & $3886.7 \pm 2.3$ & $29.5 \pm 2.4$ \\
\midrule
\multirow{4}{*}{$Y(4220)$}
 & This work & $4223.6 \pm 0.9$ & $37.7 \pm 1.7$ \\
 & BESIII PWA
 & $4225.7 \pm 4.1 \pm 3.4$ & $57.5 \pm 9.4 \pm 12.1$ \\
 & BESIII total
 & $4221.4 \pm 1.5 \pm 2.0$ & $41.8 \pm 2.9 \pm 2.7$ \\
 & PDG $\psi(4230)$
 & $4221.7 \pm 2.5$ & $51 \pm 7$ \\
\midrule
\multirow{3}{*}{$Y(4320)$}
 & This work & $4283 \pm 4$ & $102 \pm 8$ \\
 & BESIII total
 & $4298 \pm 12 \pm 26$ & $127 \pm 17 \pm 10$ \\
 & PDG $\psi(4360)$
 & $4373 \pm 7$ & $124 \pm 13$ \\
\bottomrule
\end{tabular}
\caption{
Masses and widths of the fitted resonant states compared with the BESIII PWA \cite{BESIII:2025qkn}, BESIII analysis of the total cross section $\sigma(e^+e^-\to\pi^+\pi^-J/\psi)$ \cite{BESIII:2022qal}, and with the PDG values \cite{ParticleDataGroup:2026mpi}. For the BESIII results, the first uncertainty is statistical and the second is systematic. Our fit results include only $2\sigma$ uncertainties from the experimental data.}
\label{tab:resonance-comparison}
\end{table}


The fitted $Z_c^\pm(3900)$ mass and width are consistent with the BESIII partial wave analysis (PWA) result \cite{BESIII:2025qkn}. The $Y(4220)$ parameters are also compatible with the BESIII determinations from the subprocess \cite{BESIII:2025qkn} and total cross section analyses \cite{BESIII:2022qal}. For the second vector structure, our fitted mass is closer to the BESIII $J/\psi\pi^+\pi^-$ total cross section value \cite{BESIII:2022qal} rather than to the PDG $\psi(4360)$ average \cite{ParticleDataGroup:2026mpi}. This is not unexpected, since the latter combines information from different final states. We therefore regard the $Y(4320)$ contribution in the present fit as a phenomenological second vector structure specific to the $J/\psi\pi^+\pi^-$ channel rather than as a channel-independent determination, and we discuss it in more detail in the following section.

\subsection{Comparison with related analyses and limitations of the isobar assumption}
\label{sec:comparison}

The BESIII invariant-mass distributions analyzed here have recently been described in Ref.~\cite{Chen:2026fnz} with inclusion of open-charm diagrams accounting for triangle-singularity effects, which allows for a direct comparison on the same dataset. The $e^+e^-$ energy dependence, however, is not modeled there: the vector source coupling to the virtual photon is left unspecified. The present analysis is complementary in this respect, since the $s$ dependence is fixed globally and a single set of energy-independent parameters is required to reproduce the invariant-mass distributions and the measured total cross sections simultaneously.

The DPD construction employed here is model independent as far as the Dalitz-plot function is saturated by a coherent sum of three two-body isobar chains, each carrying a line shape $X_s(\sigma_k)$ that depends on a single Mandelstam variable. Such an amplitude does not contain the singularities associated with the cross-channel exchanges, which generate nonfactorizable topologies of the triangle and box type. As discussed in Sec.~\ref{sec2}, this appears to be numerically adequate over the energy range considered. The most direct interpretational consequence of the employed model concerns the number of vector structures. Apart from the kinematic factors of the DPD construction, the $s$ dependence of the production amplitude enters exclusively through the two Breit-Wigner propagators $P_{Y_{1,2}}(s)$, the nonresonant term of Eq.~\eqref{eq:PiB} being independent of $s$. The pronounced asymmetry of the measured $J/\psi\pi^+\pi^-$ line shape must consequently be generated by the interference of two poles, and this is what drives the need for the $Y(4320)$ term in the total fit. Analyses that retain the open-charm loops have an alternative mechanism available: in Ref.~\cite{vonDetten:2024eie} the box contributions enhance the cross section at the $D_1\bar{D}$ threshold and, combined with the interference with the $\psi(4160)$, reproduce the same asymmetry with a single vector pole. To quantify how strongly the present data require the second structure within this framework, we have repeated the total fit with the $Y(4320)$ couplings set to zero, retaining the $f_2(1270)$ contribution. The description becomes less optimal, with $\chi^2/N_{\text{d.o.f.}}=1.66$ shifting to $2.55$. It is nevertheless notable that the resulting single-$Y$ fit returns $M_{Y_1}=4227.0 \pm 0.6$ and $\Gamma_{Y_1}=48.7 \pm 1.3~\mathrm{MeV}$, in close agreement with the pole position $\sqrt{s_{\text{pole}}}=(4227 \pm 4)-\frac{i}{2}(50^{+8}_{-2})~\mathrm{MeV}$ obtained in Ref.~\cite{vonDetten:2024eie} within a very different dynamical framework. The two descriptions therefore agree on the location of the leading vector structure and differ on whether the remaining energy dependence is attributed to an additional pole or to open-charm dynamics.

At the same time, the inclusion of open-charm dynamics does not remove the second structure universally ~\cite{Dong:2026fdi, Nakamura:2023obk}. The entries in Table~\ref{tab:resonance-comparison} are to be compared with $4308.1 \pm 2.2$ and $138.2 \pm 4.4~\mathrm{MeV}$ from Ref.~\cite{Nakamura:2023obk} and with the pair of dynamically generated poles at $(4278\pm3)-i\,(97\pm3)$ and $(4302\pm2)-i\,(69\pm2)~\mathrm{MeV}$ found in Ref.~\cite{Dong:2026fdi} in the model without bare seed states. What separates these descriptions is the amount of energy dependence carried by the production mechanism itself, which cannot be fixed by the $J/\psi\pi^+\pi^-$ data alone. We accordingly regard the $Y(4320)$ extracted here as an effective, channel-specific pole of the present parametrization. Resolving the $Y$ line shapes without deciding in advance whether the additional energy dependence is to be assigned to a pole or to a kinematic effect is left for a future work.

\section{Summary and conclusions}
\label{sec4}

We have presented a simultaneous analysis of $e^+e^- \to J/\psi\pi^+\pi^-$ and $e^+e^- \to J/\psi K^+K^-$ in the energy range $4.13 \le \sqrt{s} \,(\mathrm{GeV}) \le 4.36$ using the Dalitz-plot decomposition formalism. The $e^+e^-$ energy dependence is incorporated through the $Y(4220)$ and $Y(4320)$ vector structures, together with a nonresonant production term, while the scalar $\pi\pi/K\bar K$ final-state interaction is treated dispersively using a coupled-channel Omn\`es representation. 

We find that a purely resonant description is not sufficient. A nonresonant scalar production term followed by $\pi\pi/K\bar K$ rescattering is required to describe the total cross sections and invariant-mass distributions consistently. In the full fit, the inclusion of the second vector structure and the $f_2(1270)$ contribution improves the description over the full energy range, and reduces the size of the effective scalar background.

Within the present isobar model, we extract effective Breit-Wigner parameters for the $Z_c(3900)$, $Y(4220)$, and $Y(4320)$. The $Z_c(3900)$ and $Y(4220)$ parameters are compatible with the corresponding BESIII determinations, while the fitted $Y(4320)$ mass is closer to the BESIII $J/\psi\pi^+\pi^-$ result than to the PDG $\psi(4360)$ average.  The framework also allows us to extract the subprocess cross sections for $\pi^\pm Z_c^\mp(3900)$ production and for the scalar $J/\psi(\pi^+\pi^-)_{S\text{-wave}}$ channel, as well as to provide extrapolations for invariant-mass distributions at energies where such measurements are not yet available. 

In a strictly isobar-based construction, the pronounced asymmetry of the $J/\psi\pi^+\pi^-$ line shape can only be generated by the interference of two poles. Approaches that include open-charm loop mechanisms provide an alternative source of energy dependence in the same region ~\cite{vonDetten:2024eie, Chen:2026fnz, Dong:2026fdi, Nakamura:2023obk}.

\section*{Acknowledgments}
This work was supported by the Deutsche Forschungsgemeinschaft (DFG, German Research Foundation) within the Research Unit [Photon-photon interactions in the Standard Model and beyond, Projektnummer 458854507 - FOR 5327].

\bibliographystyle{apsrev4-1}
\bibliography{bib}

\end{document}